\documentclass[11pt]{article}
\usepackage{mydef2col}
\usepackage{kantlipsum,widetext}
\usepackage[T1]{fontenc}
\usepackage{array}
\usepackage[autostyle=true]{csquotes}

\allowdisplaybreaks[3]
\usepackage{tikz}
\usetikzlibrary{calc,decorations.markings}
\usetikzlibrary{arrows,patterns,positioning}

\usepackage[round]{natbib}

\def\calL{{\cal L}}

\def\baru{{\bar{u}}}
\def\barg{{\bar{\gamma}}}

\def\Jnu{{J_{|\nu|}}}
\def\Ynu{{Y_{|\nu|}}}
\def\Inu{{I_{|\nu|}}}
\def\Knu{{K_{|\nu|}}}

\def\hz{{\hat z}}
\def\pade{Pad{\'e} }

\newcommand{\iu}{\mathrm{i}\mkern1mu}

\DeclareMathOperator{\erfc}{erfc}

\title{Semi-closed form prices of barrier options  in the time-dependent CEV and CIR models}
\def\thetitle1{Semi-closed form solutions for barrier options ...}
\author{
\authorstyle{
Peter Carr,
\textsuperscript{1}
Andrey Itkin{}
\textsuperscript{1}
and Dmitry Muravey
\textsuperscript{2}
}
\newline\newline
\textsuperscript{1}
\institution{Tandon School of Engineering, New York University, 1 Metro Tech Center, 10th floor, Brooklyn NY 11201, USA} \\
\textsuperscript{2}
\institution{Moscow State University, Moscow, Russia}
}

\date{\today}
\begin{document}

\maketitle

\lettrineabstract{We continue a series of papers where prices of the barrier options written on the underlying, which dynamics follows some one factor stochastic model with time-dependent coefficients and the barrier, are obtained in semi-closed form, see (Carr and Itkin, 2020, Itkin and Muravey, 2020). This paper extends this methodology to the CIR model for zero-coupon bonds, and to the CEV model for stocks which are used as the corresponding underlying for the barrier options. We describe two approaches. One is generalization of the method of heat potentials for the heat equation to the Bessel process, so we call it the method of Bessel potentials. We also propose a general scheme how to construct the potential method for any linear differential operator with time-independent coefficients. The second one is the method of generalized integral transform, which is also extended to the Bessel process. In all cases, a semi-closed solution means that first, we need to solve numerically a linear Volterra equation of the second kind, and then  the option price is represented as a one-dimensional integral. We demonstrate that computationally our method is more efficient than both the backward and forward finite difference methods while providing better accuracy and stability.
Also, it is shown that both method don't duplicate but rather compliment each other, as one provides very accurate results at small maturities, and the other one - at high maturities.}

\vspace{0.5in}

\section*{Introduction}

This paper continues a series of papers where prices of the barrier options written on the underlying, which dynamics follows some one factor stochastic model with time-dependent coefficients and the barrier, are constructed in semi closed form, see \citep{CarrItkin2020, ItkinMuravey2020}.  Here we extend our approach for two additional models: the Cox–Ingersoll–Ross (CIR) model, \citep{cir:85}, and the time-dependent constant elasticity of variance (CEV) model, \citep{c75}. Both models are very popular among practitioners and used for pricing various derivatives such asset classes as Equities, Fixed Income, Commodities, FX, etc.

For pricing the time-dependent barrier options we develop in parallel two analytic methods, both based on the notion of generalized integral transform, \citep{CarrItkin2020, ItkinMuravey2020}. The first method is the method of heat potentials which, as applied to the models considered in this paper, is discussed in detail in Section~\ref{BP}. We extend this method, since the partial differential equation  (PDE) we need to solve cannot be transformed to the heat equation, but rather to the Bessel PDE. This approach is new and has not been developed yet in the literature. However, once this is done, the same method could be used for solving some other problems implicitly related to pricing of barrier options, e.g., pricing American options, \citep{CarrItkin2020}, analyzing the stability of a single bank and a group of banks in the structural default framework, \citep{KaushLiptonRes2018}, calculating the hitting time density, \citep{Alili2005, LiptonKau2020-2}, finding an optimal strategy for pairs trading, \citep{LiptonPrado2020}, etc. Also, the method could be used for solving various problems in physics where it was originally developed for the heat equation, \citep{kartashov2001,Friedman1964} and references therein.

The other method is the method of generalized integral transform was actively developed by the Russian mathematical school to solve parabolic equations at the domain with moving boundaries, see e.g., \citep{kartashov1999} and references therein. However, again for the Bessel PDE this approach has not been developed yet in the literature in full (i.e., up to the final formula), despite some comments on possible ways of achieving this could be found in \citep{kartashov1999}. It is also worth mentioning that so far the only known problem solved by using this method is the heat equation with the boundary $y(t)$ moving in time $t$, so the solution is defined at the domain $[0, y(t)]$. In \citep{ItkinMuravey2020} this approach was extended to the domain $[y(t),\infty)$. Also, the problem at the domain $[y(t),z(t)]$, which  emerges, eg, for the time-dependent double barrier options where the underlying follows a time-dependent Hull-White model, was also constructed in \citep{ItkinMuravey2020}.

Going back to the CIR and CEV models with time-dependent coefficients, the prices of the barrier options in these models are not known in closed form yet. Instead, various numerical methods are used to compute them. This obviously produces a computations burden which could be excessive when these numerical procedures are used as a part of calibration process. Therefore, our closed form solutions could be of importance for practitioners. By a semi-closed solution we mean that first, one needs to solve numerically a linear Volterra equation of the second kind, and then  the option price is represented as a one-dimensional integral of thus found solution. We demonstrate that computationally our method is more efficient than both the backward and forward finite difference methods while providing better accuracy and stability.

Overall, our contribution to the existing literature is twofold. First, we solve the problem of pricing barrier options in the CIR and CEV models in semi-closed form, and provide the resulting expressions not known yet in the literature. Second, we solve these problems by two methods which the extension of the existing methods and are developed by the authors.

The rest of the paper is organized as follows. Section~\ref{SecBO} shortly describes the CEV model and shows how to transform the pricing equation to the Bessel PDE. In Section~\ref{sCIR} we do same for the CIR model. Note, that we use the CEV model to price barrier options written on Equities, while the CIR model is used to price barrier options written on Zero-coupon bonds. Nevertheless, the transformed PDE is same for both models, while solution could be defined at same or different domains. In Section~\ref{BP} we develop a method of  Bessel potentials which is an extension of the method of heat potentials, and obtain the solution of our problems using this approach. In Section~\ref{secSol} same program is fulfilled for the method of generalized integral transform. In Section~\ref{numRes} the results of some numerical experiments are presented which compare the performance and accuracy of our method with  the finite-difference method used to solve the forward Kolmogorov equation (this is currently the standard way to price barrier options in the time-dependent CIR and CEV models). The last Section concludes.

Also, we discovered that the potential method could be constructed for any PDE where the space operator is a linear differential operator with constant coefficients. We propose and discuss this generalization of the heat potential method in Appendix.

\section{The CEV model} \label{SecBO}

The time-dependent constant elasticity of variance (CEV) model is a one-dimensional diffusion process that solves a stochastic differential equation %
\begin{equation} \label{CEV}
d S_t = \mu(t) S_t dt + \sigma(t) S_t^{\beta+1} dW_t, \qquad S_{t=0} = S_0.
\end{equation}
Here $t \ge 0$ is the time, $S_t$ is the stochastic stock price, $\mu(t)$ is the drift, $\sigma(t)$ is the volatility and $\beta$ is the elasticity parameter such that $\beta < 1, \ \beta \ne \{0, -1\}$\footnote{In case $\beta = 0$ this model is the Black-Scholes model, while for $\beta = -1$ this is the Bachelier, or time-dependent Ornstein-Uhlenbeck (OU) model.}, $W_t$ is  the standard Brownian motion. It is known, that under risk-neutral measure $\mu(t) = r(t) - q(t)$ where $r(t)$ is the deterministic short interest rate, and $q(t)$ is the continuous dividend. We assume that all parameters of the model are known either as a continuous functions of time $t \in [0,\infty)$, or as a discrete set of $N$ values for some moments $t_i, \ i=1,\ldots,N$.

The CEV model with constant coefficients has been introduced in \citep{c75} as an alternative to the geometric Brownian motion for modeling asset prices. Despite some sophistication as compared with the Black-Scholes model, the model is still analytically tractable, and prices of the European options can be obtained in closed-form. That is because the CEV process with constant coefficients is related to the Bessel process, \citep{LinetskyMendozza2010}. Also, as mentioned in that reference, the elasticity parameter $\beta$ controls the steepness of the skew (the larger the $|\beta|$ - the steeper the skew), while the volatility (scale) parameter $\sigma$ fixes the at-the-money volatility level. This ability to capture the skew has made the CEV model popular in equity options markets.

For the standard CEV process with constant parameters it is known that change  of  variable  $z_t = 1/(\sigma|\beta|)S_t^{-\beta}$ reduces the  CEV  process  without  drift  ($\mu = 0$)  to  the  standard Bessel process of order $1/2 \beta$ (see \citep{RevuzYor1999, DavidovLinetsky2001}). Then the continuous part of the risk-neutral density of $S_t$, conditional on $S_0 = S$, is obtained from the well-known expression for transition density of the Bessel process. If $\mu \ne 0$, using  the  result of \citep{Goldenberg:1991} this CEV  process  could be obtained  from  the process without drift via a scale and time change
\begin{equation*}
S_t^\mu = e^{\mu t} S_{\tau(t)}^0, \qquad  \tau(t) = \frac{1}{2 \mu \beta}\left(e^{2 \mu\beta t} -1\right).
\end{equation*}
Also, as shown in \citep{DavidovLinetsky2001}, at  $\beta > 0$ according to Feller's classification the origin $S_t = 0$ is a natural boundary, and infinity is an entrance boundary.

Below we show that a similar connection with the Bessel process can be established for the time-dependent version of the model in \eqref{CEV}. Let us consider an Upper-and-Out barrier Call option $C(t,S)$ written on the underlying process $S_t$. By the Feynman–Kac formula this price solves the following partial differential equation (PDE)
 \begin{equation} \label{PDE}
\fp{C}{t} + \dfrac{1}{2}\sigma^2(t) S^{2 \beta + 2}\sop{C}{S} +  [r(t) - q(t)] S \fp{C}{S} = r(t) C.
\end{equation}
This equation should be solved subject to the terminal condition at the option maturity $t=T$
\begin{equation} \label{tc0}
C(T,S) = (S-K)^+,
\end{equation}
\noindent where $K$ is the option strike, and the boundary conditions
\begin{equation} \label{bc0}
C(t,0) = 0, \qquad C(t,H(t)) = 0,
\end{equation}
\noindent where $H(t)$ is the upper barrier, perhaps time-dependent. Then the following Proposition holds.
\begin{proposition}
The PDE in \eqref{PDE} can be transformed to
\begin{equation} \label{Bess}
\fp{u}{\tau} =\frac{1}{2} \sop{u}{z} + \frac{b}{z} \fp{u}{z},
\end{equation}
\noindent where $b$ is some constant, $u = u(\tau, z)$ is the new dependent variable, and $(\tau, z)$ are the new independent variables.
The \eqref{Bess} is the PDE associated with the one-dimensional Bessel process, \citep{RevuzYor1999}
\begin{equation} \label{BesProc}
d X_t = d W_t  + \frac{b}{X_t} dt.
\end{equation}

\begin{proof}[{\bf Proof}]
This transformation can be done in two steps. First, we make a change of variables
\begin{equation} \label{tr1}
S = \left(-x \beta \right)^{-1/\beta}, \qquad C(t,S) \to u(t,x) e^{\int_0^t r(k) d k}, \qquad
\phi = \int_t^T \sigma^2(k) dk.
\end{equation}
This reduces the PDE in \eqref{PDE} to the form
\begin{align} \label{PDE1}
\fp{u}{\phi} &= \frac{1}{2}\sop{u}{x} + \left(x f(t) +  \frac{b}{x} \right) \fp{u}{x}, \\
f(t) &= \beta \frac{r(t) - q(t)}{\sigma^2(t)}, \qquad b = \frac{\beta + 1}{2\beta}, \qquad t = t(\phi). \nonumber
\end{align}
The function $t(\phi)$ is the inverse map of the last term in \eqref{tr1}. It can be computed for any $t \in [0,T]$ by substituting it into the definition of $\phi$, then finding the corresponding value of $\phi(t)$, and finally inverting.

The \eqref{PDE1} is a known type of PDE. Therefore, following \citep{Polyanin2002}, at the second step we make a new change of variables
\begin{equation} \label{tr2}
z = x F(\phi), \qquad \tau  = \int_0^\phi F^2(k) dk, \qquad F(\phi) = e^{\int_0^\phi f(k) dk}.
\end{equation}
After this change the final PDE takes the form of \eqref{Bess} which finalizes the proof.
\end{proof}
\end{proposition}
Note, that Carr and Linetsky  in \citep{CarrLinetsky2006} extend the time-dependent CEV model considered in this paper by allowing a jump to zero.
They also reduce their stock price process to a time homogeneous Bessel process. We will discuss this extension as applied to the barrier options further in this paper.

As far as the terminal (now the initial ) condition in \eqref{tc0} and the boundary conditions in \eqref{bc0} in the new variables is concerned,
we must distinguish two cases, which are determined by the sign of $\beta$. If $-1 < \beta < 0$, the domain of definition for $z$ is
$z \in [0, y(\tau)]$, where
\begin{equation} \label{ytau1}
y(\tau) = -\frac{1}{\beta} H^{-\beta} \left(t(\tau)\right) F(\phi(\tau)) > 0.
\end{equation}
Accordingly, the initial condition now reads
\begin{equation} \label{tc1}
u(0,z) =e^{-\int_0^T  r(k) dk }  \left[\left(- \frac{\beta  z}{F(\phi(0))}\right)^{-1/\beta } - K\right]^+,
\end{equation}
\noindent and the boundary conditions are
\begin{equation} \label{bc1}
u(\tau,0) = u(\tau,y(\tau)) = 0.
\end{equation}

However, if $0 < \beta < 1$, the left boundary goes to $-\infty$. Therefore, in this case it is convenient to redefine $x \to \bar{x} = -x$. This also redefines $z \to \bar{z} = -z$. Then the domain of definition for $\bar{z}$ becomes $\bar{z} \in [y(\tau), \infty)$ where
\begin{equation} \label{ytau12}
y(\tau) = \frac{1}{\beta} H^{-\beta}\left(t(\tau)\right) F(\phi(\tau)) > 0.
\end{equation}
The initial condition now reads
\begin{equation} \label{tc2}
u(0,\bar{z}) =e^{-\int_0^T  r(k) dk }  \left[\left(\frac{\beta  \bar{z}}{F(\phi(0))} \right)^{-1/\beta } - K\right]^+,
\end{equation}
\noindent and the boundary conditions are
\begin{equation} \label{bc2}
u(\tau,\bar{z})\Big|_{\bar{z} \to \infty} = u(\tau,y(\tau)) = 0.
\end{equation}
Also, in the case $0 < \beta < 1$ the PDE in \eqref{Bess} keeps the same form in the $\bar{z}$ variables. It can be seen, that in this case the Up-and-out barrier options transforms to the Down-and-out counterpart.

We interrupt here dealing with the time-dependent CEV model and postpone construction of the solution of problems in \eqref{Bess}, \eqref{tc1}, \eqref{bc1} (or \eqref{Bess}, \eqref{tc2}, \eqref{bc2}) till Section~\ref{BP}. Instead, in the next Section we consider the time-dependent CIR model, and barrier options written on a Zero-coupon bond (ZCB) which follows the CIR model. This is done, because, as we show below, the corresponding PDE could also be transformed to \eqref{Bess}. Thus, our proposed method could be applied uniformly to both models.

\section{The CIR model} \label{sCIR}

The Cox–Ingersoll–Ross (CIR) model has been invented in \citep{cir:85} for modeling interest rates. In this model the instantaneous interest rate $r_t$ is stochastic variable which follows the stochastic differential equation (SDE), also named the CIR process. For the time-dependent version of the model this SDE reads
\begin{equation} \label{OU1}
d r_t = \kappa(t)[\theta(t) - r_t] dt + \sigma(t)\sqrt{r_t} dW_t, \qquad r_{t=0} = r.
\end{equation}
Here $\kappa(t) > 0$ is the constant speed of mean-reversion, $\theta(t)$ is the mean-reversion level. The CIR model is an extension of the Hull-White model that we analyzed in \citep{ItkinMuravey2020} by making the volatility proportional to $\sqrt{r_t}$.  This, on the one hand, allows avoiding the possibility of negative interest rates when the Feller condition $ 2 \kappa(t) \theta(t)/\sigma^2(t) > 1$ is satisfied, while, on the other hand, still preserves tractability of the model, see e.g., \citep{andersen2010interest} and references therein.

Since the CIR model belongs to the class of exponentially affine models, the price of the ZCB $F(r,t,S)$ for this model is known in closed form. Here $S$ is the bond expiration time. It is known, that $F(r,t,S)$ under a risk-neutral measure solves a linear parabolic partial differential equation (PDE), \citep{privault2012elementary}
\begin{equation} \label{PDECIR}
\fp{F}{t} + \dfrac{1}{2}\sigma^2(t) r \sop{F}{r} + \kappa(t) [\theta(t) - r] \fp{F}{r} = r F.
\end{equation}
It should be solved subject to the terminal condition
\begin{equation} \label{termZCB}
 F(r,S,S)  = 1,
\end{equation}
\noindent and the boundary condition
\begin{equation} \label{bcZCB}
F(r,t,S)\Big|_{r \to \infty} = 0.
\end{equation}
The second boundary condition is necessary in case the Feller condition is violated, so the interest rate $r_t$ can hit zero. Otherwise, the PDE in \eqref{PDECIR} itself at $r=0$ serves as the second boundary condition.

The ZCB price can be obtained from \eqref{PDECIR} assuming that the solution of can be represented in the form
\begin{equation} \label{affSol}
F(r,t,S) = A(t,S) e^{ B(t,S) r}.
\end{equation}
Substituting this expression into \eqref{PDE} and separating the terms proportional to $r$, we obtain two equations to determine $A(t,S), B(t,S)$
\begin{align} \label{equAff}
 \fp{B(t,S)}{t} &= 1 + \kappa(t) B(t,S) - \frac{1}{2} \sigma^2(t) B^2(t,S), \\
 \fp{A(t,S)}{t} &= - A(t,S) B(t,S)  \theta (t) \kappa (t). \nonumber
\end{align}
To obey the terminal condition \eqref{termZCB}, the first PDE in \eqref{equAff} should be solved subject to the terminal condition $B(S,S) = 0$, and the second one - to  $A(S,S) = 1$.

The first equation in \eqref{equAff} is the Riccati equation. It this general form it cannot be solved analytically for arbitrary functions $\kappa(t),  \sigma(t)$, but can be efficiently solved numerically. Also, in some cases it can be solved approximately (asymptotically), see e.g., an example in \citep{CarrItkin2020}.  Once the solution is obtained, the second equation in \eqref{equAff} can be solve analytically to yield
\begin{equation}
A(t,S) = e^{- \int_S^t B(m) \theta (m) \kappa (m) \, dm}.
\end{equation}
When coefficients $\kappa(t), \theta(t), \sigma(t)$ are constants, it is known that the solution $B(t,S)$ can be obtained in closed form and reads, \citep{andersen2010interest}
\begin{equation}
B(t,S) = - \frac{2[\exp((S-t)h) - 1]}{2h + (\theta + h)[\exp((S-t)h) - 1]}, \qquad h = \sqrt{\theta^2 + 2 \sigma^2}.
\end{equation}
Thus, $B(t,S) < 0$ if $t < S$. Therefore,  $F(r,t,S) \to 0$ when $r \to \infty$. In other words, the solution in \eqref{affSol} satisfies the boundary condition at $r \to \infty$. In case when all the parameters of the model are deterministic functions of time, and $B(t,S)$ solves the first equation in \eqref{equAff}, this also remains to be true. This can be checked as follows. Since $\kappa(t) > 0, \sigma(t) > 0$ and from \eqref{affSol}
\begin{equation*}
B(t,S) = \frac{\kappa(t)}{\sigma(t)^2} - \frac{1}{\sigma^2(t)} \sqrt{\kappa(t)^2 + 2 [1 - B'(t,S)]}.
\end{equation*}
Therefore, $B(t,S) < 0$ if $B'(t,S) < 1$. On the other hand, if $B(t,S) < 0$ then from \eqref{affSol} $B'(t,S) < 1$. This finalizes the proof.

\subsection{Down-and-Out barrier option} \label{DOB}

Let us consider a Down-and-Out barrier Call option written on a ZCB. Under a risk-neutral measure the option price $C(t,r)$ solves the same PDE as in  \eqref{PDECIR}, \citep{andersen2010interest}.
\begin{equation} \label{PDEP}
\fp{C}{t} + \dfrac{1}{2}\sigma^2(t) r\sop{C}{r} + \kappa(t) [\theta(t) - r] \fp{C}{r} = r C.
\end{equation}
The terminal condition at the option maturity $T \le S$ for this PDE reads
\begin{equation} \label{tc0CIR}
C(T,r) = \left(F(r,T,S) - K\right)^+,
\end{equation}
\noindent where $K$ is the option strike.

By a standard contract, the lower barrier $L_{F}(t)$ (which we assume to be time-dependent as well) is set on the ZCB price, and not on the underlying interest rate $r$. This means that it can be written in the form
\begin{equation} \label{ZCBBar}
C(t, r) = 0 \quad \mbox{if } F(r,t,S) = L_F(t).
\end{equation}
However, since the ZCB price $F(r,t,S)$ can be expressed in closed form in \eqref{affSol}, this condition can be translated into the $r$ domain by solving the equation
\begin{equation*}
F(r,t,S) = A(t,S) e^{B(t,S) r} = L_F(t),
\end{equation*}
\noindent with respect to $r$. Denoting the solution of this equation as $L(t)$ we find
\begin{equation} \label{Loft}
L(t) =  \frac{1}{B(t,S)} \log \left(\frac{L_F(t)}{A(t,S)}\right) > 0,
\end{equation}
\noindent where it is assumed that $L_F > A(t,S)$. Accordingly, in the $r$ domain the boundary condition to \eqref{PDEP} reads
\begin{equation} \label{bcBar}
C(t,L(t)) = 0.
\end{equation}
The second boundary can be naturally set at $r \to \infty$. As at $r \to \infty$ the ZCB price tends to zero, see \eqref{affSol}, the Call option price also vanishes in this limit. This yields
\begin{equation} \label{bc01}
 C(t,r)\Big|_{r \to \infty}  = 0.
\end{equation}

The PDE in \eqref{PDEP} can also be transformed to that for the Bessel process in \eqref{BesProc}.

\begin{proposition} \label{prop2}
The \eqref{PDEP} can be transformed to
\begin{equation} \label{Bess1}
\fp{u}{\tau} = \frac{1}{2} \sop{u}{z} + \frac{b}{z} \fp{u}{z},
\end{equation}
\noindent where $b$ is some constant, $u = u(\tau, z)$ is the new dependent variable, and $(\tau, z)$ are the new independent variables, if
\begin{equation} \label{cond1}
\frac{\kappa(t) \theta(t)}{\sigma^2(t)} = \frac{m}{2},
\end{equation}
\noindent where $m \in [0,\infty)$ is some constant. The \eqref{Bess1} is the PDE associated with the one-dimensional Bessel process in \eqref{BesProc}.

\begin{proof}[{\bf Proof}] First make a change of variables

\begin{align} \label{trCIR1}
C(t,r) &= u(t,z) e^{a(t) r + \int_0^t a(s) \kappa(s) \theta(s) ds}, \qquad z = g(t) \sqrt{r}, \\
g(t) &=  \exp\left[ \frac{1}{2} \int_0^t \left( \kappa(s) - a(s) \sigma^2(s) \right) \, ds \right]
\end{align}
\noindent where $a(t)$ solves the Riccati equation
\begin{equation} \label{ric2}
\frac{da(t)}{dt} = -\frac{\sigma^2(t) a^2(t)}{2} + \kappa(t) a(t)  + 1.
\end{equation}

This reduces the PDE in \eqref{PDEP} to the form
\begin{align} \label{PDE12}
\frac{4 k(t) \theta (t) - \sigma^2(t)}{2 z} \fp{u}{z} + \frac{1}{2} \sigma (t)^2 \sop{u}{z} + \frac{4}{g^2(t)}\fp{u}{t} = 0.
\end{align}
Now make a change of time
\begin{equation} \label{tauTr}
\tau(t) = \frac{1}{4} \int_t^T  g^2(s) \sigma^2(s) \, ds,
\end{equation}
\noindent which transforms \eqref{PDE12} to
\begin{equation} \label{PDE3}
\left(2 \frac{k(t) \theta (t)}{\sigma^2(t)} - \frac{1}{2}\right) \frac{1}{z} \fp{u}{z} + \frac{1}{2}\sop{u}{z} = \fp{u}{\tau}, \qquad t = t(\tau).
\end{equation}
The function $t(\tau)$ is the inverse map of \eqref{tauTr}. It can be computed for any $t \in [0,T]$ by substituting it into the definition of $\tau$, then finding the corresponding value of $\tau(t)$, and inverting.

Finally, as by assumption $k(t) \theta (t)/\sigma^2(t) = m - const$, we set $b = m - 1/2$. Thus, the final PDE takes the form of \eqref{Bess} which finalizes the proof.
\end{proof}
\end{proposition}

As follows from Proposition ~\ref{prop2}, for the time-dependent CIR model the transformation from \eqref{PDEP} ro \eqref{Bess1} cannot be done unconditionally. However, from practitioners' points of view the condition \eqref{cond1} seems not to be too restrictive, Indeed, the model parameters already contain the independent mean-reversion rate $\kappa(t)$ and volatility $\sigma(t)$.  Since $m$ is an arbitrary constant, it could be calibrated to the market data together with $\kappa(t)$ and $\sigma(t)$. Therefore, in this form the model should be capable for calibration to the term-structure of interest rates.

Also, according to the change of variables made in Proposition ~\ref{prop2}, the terminal condition \eqref{tc0CIR} in new variables reads
\begin{equation} \label{tc0CIRu}
u(0,z) = e^{-a(t(0)) z^2 / g^2(t(0)) - \int_0^T a(s)\kappa(s)\theta(s) ds}\left( A(T,S) e^{B(T,S) z^2/g^2(T)} - K\right)^+.
\end{equation}
And the boundary conditions in \eqref{bcBar} and \eqref{bc01} transform to
\begin{align} \label{bcCIRu}
C(\tau,y(\tau)) &= 0, \qquad C(\tau,z)\Big|_{z \to \infty}  = 0, \\
y(\tau) &= \frac{1}{B(t(\tau),S(\tau))} \log \left(\frac{L_F(t(\tau))}{A(t(\tau),S(\tau))}\right). \nonumber
\end{align}

\section{The method of Bessel potentials} \label{BP}
For convenience of notation, further let us call as {\it the CEV problem} the PDE in \eqref{Bess} that has to be solved subject to the initial condition in \eqref{tc2} and the boundary conditions in \eqref{bc2}. Also, we call {\it the CIR problem} the PDE in \eqref{Bess1} solved subject to the initial condition \eqref{tc0CIRu} and the boundary conditions in \eqref{bcCIRu}. Both problems can be considered simultaneously, as the PDE in \eqref{Bess} coincides with that in \eqref{Bess1}, and the boundary conditions in \eqref{bc2} coincide with that in \eqref{bcCIRu}. Accordingly, both solutions are defined at the domain $z \in [y(\tau), \infty]$, thought the  definitions of $y(\tau)$ in \eqref{ytau12} and \eqref{bcCIRu} differ.
We will describe our method for this domain in Section~\ref{infDom}. Another type of the CEV problem where $\bar{z} \in [0,y^-(\tau)]$ will be considered separately in Section~\ref{fixDom}.

In \citep{CarrItkin2020, ItkinMuravey2020} a similar problem but for the heat equation was solved by using two approaches. The first one is
a method of generalized integral transform, actively elaborated on by the Russian mathematical school to solve parabolic equations at the domain with moving boundaries, see \citep{kartashov1999,kartashov2001} and references therein. These kind of problems are known in physics for a long time and arise in the field of nuclear power engineering and safety of nuclear reactors; in studying combustion in solid-propellant rocket engines; in the theory of phase transitions (the  Stefan  problem  and  the  Verigin  problem  (in  hydromechanics)); in the processes of sublimation in freezing and melting; in  the  kinetic  theory  of  crystal  growth; etc. In \citep{CarrItkin2020,kartashov1999} this method was applied to the domain $z \in [0,y(\tau)]$, and in \citep{ItkinMuravey2020} for the first time the solution was obtained for the semi-infinite domain $z \in [y(\tau),\infty)$.
We will further develop this method to solve the CEV and CIR problems in Section~\ref{secSol}.

The second method used to solve the same problems in \citep{CarrItkin2020, ItkinMuravey2020} is the method of heat potentials, see, e.g., \citep{TS1963, Friedman1964, kartashov2001} and references therein. The first use of this method in mathematical finance is due to \citep{Lipton2002} for pricing path-dependent options with curvilinear barriers, and more recently in \citep{ LiptonPrado2020} (also see references therein). However, the CEV and CIR problems which we deal with in this paper, cannot be reduced to the heat equation, but rather to the Bessel PDE. Therefore, in this Section we propose generalization of the method for this type of equations. Accordingly, we call this generalization as the method of Bessel potentials.

Note, that the potential method could be constructed for any PDE where the space operator is a linear differential operator with time-independent  coefficients. We propose and discuss this generalization of the heat potential method in Appendix.  Thus, the heat and Bessel potentials are just two particular cases of this general scheme.

\subsection{Domain $y(\tau) \le z < \infty$.} \label{infDom}

Since both the CEV and CIR problems have the inhomogeneous initial condition, our first step is to reduce them to the alternative problems with a homogeneous  initial condition. Since the Green function of the Bessel equation at the infinite domain is known in closed form, \citep{Polyanin2002}, this can be achieved by representing  $u(\tau,z)$ in the form
\begin{equation} \label{q}
u(\tau,z) = q(\tau,z) + \int_{y(0)}^\infty u(0,\xi) q_\tau(z,\zeta,b) d \zeta.
\end{equation}
Here $q_\tau(z,\zeta,b)$ is the fundamental solution (or the transition density, or the Green function) of \eqref{Bess1} at the domain $z \in [0,\infty)$. This density can be obtained assumed that the Bessel process stops when it reaches the origin.  But since the domain of definition of $z$ is $z \in [y(\tau),\infty)$, we moved the left boundary from $0$ to $y(0)$.

By the definition of $b$ in \eqref{PDE1}, for the CEV model $b =  (1 + \beta)/(2\beta)$. Since in this case $0 < \beta < 1$, we get
$b > 1$. It is known, \citep{Lawler2018NotesOT,LinetskyMendozza2010},  that in case $b \ge 1/2$ the density $q_\tau(z,\zeta,b)$ is a good density with no defect of mass, i.e., it integrates into 1. The explicit representation reads, \citep{c75, Emanuel:1982}
\begin{equation} \label{Green}
q_\tau(z,\zeta,b) = \frac{\sqrt{z \zeta}}{\tau} \left(\frac{\zeta}{z}\right)^b  e^{- \frac{z^2 + \zeta^2}{2 \tau}} I_{b-1/2} \left( \frac{z \zeta}{\tau}\right).
\end{equation}
Here $I_\nu(x)$ is the modified Bessel function of the first kind, \citep{as64}.

For the CIR model $b = m - 1/2$ where $m \in [0,\infty)$ can be found by calibration. Therefore, if $m > 1$ (i.e., if the Feller condition is satisfied, and the process never hits the origin), the Green function $q_\tau(z,\zeta,b)$ is given by \eqref{Green}. Otherwise, $0 < m < 1$ and $-1/2 < b < 1/2$. Then by another change of variables, \citep{Polyanin2002}
\begin{equation*}
w(\tau,z) = z^{2(1-m)} u(\tau,z),
\end{equation*}
\noindent the \eqref{Bess1} transforms to the same equation with respect to $w(\tau,z)$ but now with $b = (3-2m)/2$. Accordingly, since $0 < m < 1$ we have $b > 1/2$, Therefore, again the Green function is represented by \eqref{Green}.

The function $q(x,\tau)$ solves the problem
\begin{align} \label{qP}
\fp{q(\tau,z)}{\tau} &= \frac{1}{2}\sop{q(\tau,z)}{z} + \frac{b}{z} \fp{q(\tau,z)}{z}, \\
q(0,z) &= 0, \qquad y(0) < z < \infty, \nonumber \\
q(\tau,z)\Big|_{z \to \infty} &= 0, \qquad q(\tau, y(\tau)) = \varsigma(\tau), \nonumber \\
\varsigma(\tau) &= - \int_{y(0)}^\infty u(0,\zeta)  q_\tau(y(\tau),\zeta,b) d \zeta. \nonumber
\end{align}
This problem is like that in \eqref{Bess1}, \eqref{tc2}, \eqref{bc2}  \eqref{tc0CIRu}, \eqref{bcCIRu}), but now with a homogeneous initial condition. Therefore, following the general idea of the method of heat potentials,  we represent the solution in the form of a generalized potential for the Bessel PDE
\begin{equation} \label{poten}
q(\tau,z) = \int_0^\tau \Psi(k) \fp{}{\xi}\left[
\frac{\sqrt{z \xi}}{\tau-k}   \left(\frac{\xi}{z}\right)^b e^{- \frac{z^2 + \xi^2}{2 (\tau-k)}} I_{b-1/2}
\left( \frac{z \xi}{\tau-k}\right) \right]\Bigg|_{\xi \to y(k)} dk,
\end{equation}
\noindent  where $\Psi(k)$ is the potential density. It can be seen that $q(\tau,z)$ solves \eqref{Bess1} as the derivative of the integral on the upper limit is proportional to the Delta function which vanishes due to $z \ne y(\tau)$. The solution in \eqref{poten} also satisfies the initial condition at $\tau = 0$, and the vanishing condition at $z \to \infty$.
For the large values of argument $z \xi /(\tau-k)$ we propose to use the following approximation :
\begin{equation}
    \label{poten_approx}
q(\tau,z) \approx \frac{1}{\sqrt{2\pi}} \int_0^\tau \Psi(k) \fp{}{\xi}\left[
\frac{1}{\sqrt{\tau-k}} \left(\frac{\xi}{z}\right)^b e^{- \frac{(z- \xi)^2}{2 (\tau-k)}} \right]\Bigg|_{\xi \to y(k)} dk
\end{equation}

At the barrier $z = y(\tau)$ function $q(\tau,z)$ is discontinuous. Following a similar approach for the heat potentials method, \citep{TS1963}),
it can be shown that the limiting value of $q(\tau,z)$ at $z = y(\tau) + 0$ is equal to $\varsigma(\tau)$:
\begin{equation} \label{Abel2k}
\varsigma(\tau) = \Psi(\tau) + \int_0^\tau \Psi(k) \fp{}{y(k)}\left[
\frac{\sqrt{y(\tau) y(k)}}{\tau-k}   \left(\frac{y(k)}{y(\tau)}\right)^b e^{- \frac{y^2(\tau) + y^2(k)}{2 (\tau-k)}} I_{b-1/2}
\left( \frac{y(\tau) y(k)}{\tau-k}\right) \right] dk.
\end{equation}
The \eqref{Abel2k} is a linear Volterra equations of the second kind, \citep{polyanin2008handbook}. Since $\varsigma(\tau)$ is a continuously differentiable function, \eqref{Abel2k} has a unique continuous solution for $\Psi(\tau)$. The Volterra equation can be efficiently solved numerically, see \citep{ItkinMuravey2020} for the discussion on various approached to the numerical solution of this type of equations. In brief, for instance, the integral in the RHS is approximated using some quadrature rule with $N$ nodes in $k$ space, and the solution is obtained at $N$ nodes in the $\tau$ space. Thus, obtained matrix equation can be solved with the complexity $O(N^2)$ since the matrix is lower triangular. Since $N$ could be small ($N \approx 20-30$), the solution is fast. In more detail we discuss numerical aspects of the solution in Section~\ref{numRes}.

Once \eqref{Abel2k} is solved and the function $\Psi(\tau)$ is found, the final solution reads
\begin{align} \label{qSol}
u(\tau,z) &=  \int_0^\tau \Psi(k) \fp{}{y(k)}\left[
\frac{\sqrt{z y(k)}}{\tau-k}   \left(\frac{y(k)}{z}\right)^b e^{- \frac{z^2 + y^2(k)}{2 (\tau-k)}} I_{b-1/2}
\left( \frac{z y(k)}{\tau-k}\right) \right] dk
+ \int_{y(0)}^\infty u(0,\xi) q_\tau(z,\zeta,b) d \zeta.
\end{align}

\subsection{Domain $0 < z < y(\tau)$.} \label{fixDom}

The construction of the solution in this case is similar to that described in the previous Section. Again, to obtain a PDE with a homogeneous initial condition, we represent the solution in the form
\begin{align} \label{q0}
u(\tau,z) &= q(\tau,z) - \varsigma_0(\tau) + \int_0^{y(0)} u(0,\xi) q_\tau(z,\zeta,b) d \zeta, \\
\varsigma_0(\tau) &= - \int_0^{y(0)} u(0,\zeta)  q_\tau(0,\zeta,b) d \zeta, \nonumber
\end{align}
\noindent where, \citep{as64}
\begin{equation*}
q_\tau(0,\zeta,b) = \frac{2^{1/2-b} \zeta^{2 b} }{\tau ^{b+1/2} \Gamma \left(b+\frac{1}{2}\right)}e^{-\frac{\zeta ^2}{2 \tau }},
\end{equation*}
\noindent and $\Gamma(x)$ is the Euler Gamma function.

Then the function $q(x,\tau)$ solves the problem
\begin{align} \label{qP0}
\fp{q(\tau,z)}{\tau} &= \frac{1}{2}\sop{q(\tau,z)}{z} + \frac{b}{z} \fp{q(\tau,z)}{z}, \\
q(0,z) &= 0, \qquad 0 < z < y(0), \nonumber \\
q(\tau,0) &= 0, \qquad q(\tau, y(\tau)) = \varsigma(\tau) + \varsigma_0(\tau). \nonumber
\end{align}

We search for the solution in the form of the Bessel potential in \eqref{poten}. The potential density $\Psi(\tau)$ solves the following Volterra equation of the second kind
\begin{equation} \label{Abel2k0}
\varsigma(\tau) + \varsigma_0(\tau)  = \Psi(\tau) + \int_0^\tau \Psi(k) \fp{}{y(k)}\left[
\frac{\sqrt{y(\tau) y(k)}}{\tau-k}   \left(\frac{y(k)}{y(\tau)}\right)^b e^{- \frac{y^2(\tau) + y^2(k)}{2 (\tau-k)}} I_{b-1/2}
\left( \frac{y(\tau) y(k)}{\tau-k}\right) \right] dk.
\end{equation}
Once this function is found, the final solution reads
\begin{align} \label{qSol0}
u(\tau,z) &=  \int_0^\tau \Psi(k) \fp{}{y(k)}\left[ \frac{\sqrt{z y(k)}}{\tau-k}   \left(\frac{y(k)}{z}\right)^b e^{- \frac{z^2 + y^2(k)}{2 (\tau-k)}} I_{b-1/2} \left( \frac{z y(k)}{\tau-k}\right) \right] dk
+ \int_0^{y(0)} u(0,\xi) q_\tau(z,\zeta,b) d \zeta.
\end{align}

\subsection{Double barrier options} \label{DBO}

Double barrier options for time-dependent  models can be also priced by using the method of potentials. In particular, in \citep{ItkinMuravey2020} this is demonstrated for the time-dependent Hull-White model. Here we use a similar approach and apply the idea proposed in \citep{ItkinMuravey2020} to construction of the semi-closed form solutions for double barrier options for the CIR and CEV models.

Let us provide the explicit formulae just for the CEV model as for the CIR model this can be done in the exact same way. Suppose we
need the price of  a double barrier Call option with the lower barrier $L(t)$ and the upper barrier $H(t) > L(t)$. After doing transformation to the Bessel PDE as this is described in Section~\ref{SecBO}, this implies solving the following problem

\begin{align} \label{uDB}
\fp{u}{\tau} &=\frac{1}{2} \sop{u}{z} + \frac{b}{z} \fp{u}{z}, \\
u(\tau=0,z) &= u(0,z), \qquad y(0) < x < h(0), \nonumber \\
u(y(\tau),\tau) &= u(h(\tau), \tau) = 0, \nonumber
\end{align}
\noindent where for $-1 < \beta < 0$
\begin{equation}
y(\tau) = -\frac{1}{\beta} L\left(t(\tau)\right)^{-\beta} F(\phi(\tau)), \qquad
h(\tau) = -\frac{1}{\beta} H\left(t(\tau)\right)^{-\beta} F(\phi(\tau)),
\end{equation}
\noindent and for $0 < \beta < 1$
\begin{equation}
y(\tau) = \frac{1}{\beta} L\left(t(\tau)\right)^{-\beta} F(\phi(\tau)), \qquad
h(\tau) = \frac{1}{\beta} H\left(t(\tau)\right)^{-\beta} F(\phi(\tau)).
\end{equation}
Thus, in this case the solution is defined at the $z$-domain with two moving (time-dependent) boundaries.

Since this problem has an inhomogeneous initial condition, the method of potentials cannot be directly applied. Therefore, similar to \eqref{q} we represent the solution in the form
\begin{equation} \label{q1}
u(\tau,z) = q(\tau,z) + \int_{y(0)}^{h(0)} u(0,\xi) q_\tau(z,\zeta,b) d \zeta.
\end{equation}
Now the function $q(x,\tau)$ solves a similar problem but with the homogeneous initial condition
\begin{align} \label{qDB}
\fp{q}{\tau} &=\frac{1}{2} \sop{q}{z} + \frac{b}{z} \fp{q}{z}, \\
q(0,z) &= 0, \qquad y(0) < x < h(0), \nonumber \\
q(\tau,y(\tau) &= - \phi_2(\tau), \qquad q(\tau, h(\tau))  = -\psi_2(\tau), \nonumber \\
\phi_2(\tau) &= - \int_{y(0)}^{h(0)} u(0,\xi) q_\tau(y(\tau),\zeta,b) d \zeta, \qquad
\psi_2(\tau) = - \int_{y(0)}^{h(0)} u(0,\xi) q_\tau(h(\tau),\zeta,b) d \zeta,. \nonumber
\end{align}

Based on the method of \citep{ItkinMuravey2020}, we construct the solution of \eqref{qDB} in the form of a generalized Bessel potential
\begin{align} \label{poten1}
q(\tau,z) = \int_0^\tau \Bigg\{ &\Psi(k) \fp{}{\xi}\left[\frac{\sqrt{z \xi}}{\tau-k}   \left(\frac{\xi}{z}\right)^b e^{- \frac{z^2 + \xi^2}{2 (\tau-k)}} I_{b-1/2} \left( \frac{z \xi}{\tau-k}\right) \right]\Bigg|_{\xi \to y(k)}  \\
+ &\Phi(k) \fp{}{\xi}\left[\frac{\sqrt{z \xi}}{\tau-k}   \left(\frac{\xi}{z}\right)^b e^{- \frac{z^2 + \xi^2}{2 (\tau-k)}} I_{b-1/2} \left( \frac{z \xi}{\tau-k}\right) \right]\Bigg|_{\xi \to h(k)}  \Bigg\} dk. \nonumber
\end{align}
Here $\Psi(k), \Phi(k)$ are the Bessel potential densities to be determined. Using the boundary conditions in \eqref{qDB}, and the fact that the expression in square brackets at $\tau = k$ is the Dirac delta function, one can find that they solve a system of two Volterra equations of the second kind
\begin{alignat}{1}  \label{Abel2k1}
\phi_2(\tau) = \Psi(\tau) + \int_0^\tau &\Bigg\{ \Psi(k) \fp{}{\xi}\left[\frac{\sqrt{y(\tau) \xi}}{\tau-k}   \left(\frac{\xi}{y(\tau)}\right)^b e^{- \frac{y^2(\tau) + \xi^2}{2 (\tau-k)}} I_{b-1/2} \left( \frac{y(\tau) \xi}{\tau-k}\right) \right]\Bigg|_{\xi \to y(k)}  \\
 + &\Phi(k) \fp{}{\xi}\left[\frac{\sqrt{y(\tau) \xi}}{\tau-k}   \left(\frac{\xi}{y(\tau)}\right)^b e^{- \frac{y^2(\tau) + \xi^2}{2 (\tau-k)}} I_{b-1/2} \left( \frac{y(\tau) \xi}{\tau-k}\right) \right]\Bigg|_{\xi \to h(k)}  \Bigg\} dk. \nonumber \\
\psi_2(\tau) = \Phi(\tau)  + \int_0^\tau \Bigg\{ &\Psi(k) \fp{}{\xi}\left[\frac{\sqrt{h(\tau) \xi}}{\tau-k}   \left(\frac{\xi}{h(\tau)}\right)^b e^{- \frac{h^2(\tau) + \xi^2}{2 (\tau-k)}} I_{b-1/2} \left( \frac{h(\tau) \xi}{\tau-k}\right) \right]\Bigg|_{\xi \to y(k)}  \\
+ &\Phi(k) \fp{}{\xi}\left[\frac{\sqrt{h(\tau) \xi}}{\tau-k}   \left(\frac{\xi}{h(\tau)}\right)^b e^{- \frac{h^2(\tau) + \xi^2}{2 (\tau-k)}} I_{b-1/2} \left( \frac{h(\tau) \xi}{\tau-k}\right) \right]\Bigg|_{\xi \to h(k)}  \Bigg\} dk. \nonumber
\end{alignat}
This system can be solved by various numerical methods with complexity $O(N^2)$ (see the discussion after \eqref{Abel2k}). Once this is done, the solution of the double barrier problem is found.

\section{The method of generalized integral transform} \label{secSol}

In this Section we solve the same problem but using the method of generalized integral transform. As applied to finance this method was successfully used in \citep{CarrItkin2020, ItkinMuravey2020} to price barrier options in the time-dependent OU model and American option for equites, and the Hull-White models for interest rates.  The method was borrowed from physics, where it was used to solve the Stefan problem and other heat and mass transfer problems with a moving boundary (or moving interphase boundary), see \citep{kartashov1999, kartashov2001} and references therein. In particular, in \citep{ItkinMuravey2020} the authors extended this approach to an infinite domain where the solution was not known yet. Below we extend this approach and apply it to the CEV and CIR problems.

Note, that so far, the method was elaborated on and used just for getting a semi closed form solution of the heat equation. But in this paper we deal with the Bessel PDE. It \citep{kartashov1999} it is proposed to construct the direct integral transform for this equation by using Bessel functions. However, except this recommendation the explicit solution has not been presented. Moreover, our analysis showed that using the form of the transform proposed in \citep{kartashov1999}doesn't give rise to the solution, as construction of the inverse transform faces various technical problems.

Therefore, our method presented in this Section is i) completely original and ii) gives rise to the closed-form solution of the problem. In other words, we solve the CIR and CEV problems by using the method of generalized integral transform to the very end, and to the best of the authors knowledge this is done for the first time in the literature. As such, this approach could be also very useful in physics for solving various problems. As mentioned in \citep{kartashov1999} those problems appear (but not limited to) in the field of nuclear power engineering and safety of nuclear reactors; in studying combustion in solid-propellant rocket engines; in  laser  action  on  solids; in the theory of phase transitions (the  Stefan  problem  and  the  Verigin  problem  (in  hydromechanics)); in the processes of sublimation in freezing and melting; in  the  kinetic  theory  of  crystal  growth; etc., see \citep{kartashov1999} and references therein.

\subsection{Domain $0 < z < y(\tau)$} \label{fixDomGIT}

Recall, that based on the description in Section~\ref{SecBO}, this problem emerges when we consider the CEV problem with $\beta < 0$.
Since the Laplace transform of \eqref{Bess1} gives rise to the Bessel ODE, \citep{as64}, it would be natural seeking for the general integral transform in the class of Bessel functions. Therefore, by analogy with \citep{kartashov2001, CarrItkin2020} we introduce the generalized integral transform of the form
\begin{equation} \label{GITdef}
\baru(\tau, p) = \int_0^{y(\tau)} z^{\nu + 1} u(\tau, z) \Jnu(z p) dz,
\end{equation}
\noindent where $ p = a + \iu \omega$ is a complex number, $J_\nu(x)$ is the Bessel function of the first kind, and
$\nu = 1/(2\beta) < 0$, since $\beta < 0$. Next, let us multiply both parts of \eqref{Bess1} by $z^{\nu + 1}\Jnu(z p)$ and integrate on $z$ from $0$ to $y(\tau)$. For the LHS this yields
\begin{align} \label{left}
\int_0^{y(\tau)} z^{\nu + 1} \frac{\partial u}{\partial \tau} \Jnu(z p) dz = \frac{\partial \baru}{\partial \tau} - y'(\tau) [y(\tau)]^{\nu + 1} u(\tau, y(\tau))  \Jnu(y(\tau) p).
\end{align}
The last term in the RHS of \eqref{left} vanishes due to the boundary condition in \eqref{bc1}.

Accordingly, for the RHS of \eqref{Bess1} we have $b = \nu + 1/2$, and
\begin{align} \label{right}
J_1 &= \frac{1}{2} \int_0^{y(\tau)}  z^{\nu+1} \sop{u}{z} \Jnu(z p) dz =
\frac{1}{2} z^{\nu+1} \fp{u}{z} \Jnu(z p)\Bigg|_0^{y(\tau)}
- \frac{1}{2} u(\tau,z)  \fp{}{z} \left(z^{\nu+1} \Jnu(z p)\right)\Bigg|_0^{y(\tau)} \\
&+ \frac{1}{2} \int_0^{y(\tau)} u(\tau,z) \sop{}{z} \left(z^{\nu+1} \Jnu(z p) \right) dz, \nonumber \\
J_2 &= \int_0^{y(\tau)}  z^{\nu+1} \frac{\nu + 1/2}{z} \fp{u}{z} \Jnu(z p) dz =
(\nu +1/2)  z^{\nu} \Jnu(z p) u(\tau,z)\Bigg|_0^{y(\tau)}  \nonumber \\
&- (\nu +1/2) \int_0^{y(\tau)}  u(\tau,z)  \fp{}{z} \left(z^\nu \Jnu(z p) \right) dz. \nonumber
\end{align}

Due to the boundary conditions in \eqref{bc1}, the sum $J_1 + J_2$ can be represented as
\begin{align} \label{sumJ}
J_1 + J_2 = y^{\nu+1}(\tau)  \Jnu(y(\tau) p) \Psi(\tau) &+
\frac{1}{2} \int_0^{y(\tau)} u(\tau,z) z \left[ \frac{1-2 \nu}{z} \fp{}{z} \left(z^\nu \Jnu(z p) \right) +
\sop{}{z}\left(z^\nu \Jnu(z p) \right) \right] dz, \nonumber \\
\Psi(\tau) &= \fp{u}{z}\Big|_{z = y(\tau)}.
\end{align}

It can be checked by using the theory of cylinder functions, \citep{bateman1953higher} that the Bessel function $\Jnu(z p)$ also solves the following ordinary differential equation (ODE)
\begin{equation}
\label{Bessel_def}
\frac{d^2 Z}{dz^2} + \frac{1-2\nu}{z} \frac{d Z}{dz}  +p^2 Z = 0,  \quad Z(z) = C_1 z^\nu \Jnu(z p) + C_2 z^{\nu}\Ynu (pz).
\end{equation}
Here $\Ynu(z)$ denotes the Bessel function of the second kind (also known as the Neumann or Weber function) which is linearly independent of $\Jnu(z)$. Assuming $C_1 = 1, \ C_2 = 0$, from \eqref{left}, \eqref{right} we obtain the following Cauchy problem for $\baru$
\begin{align} \label{baruCauchy}
\frac{d \baru(\tau,p)}{d \tau} &=  \frac{1}{2}\left[ - p^2 \baru(\tau,p) + y^{\nu + 1}(\tau) \Jnu(y(\tau) p) \Psi(\tau) \right], \\
\baru(p,0) &= \int_0^{y(0)} z^{\nu + 1}\Jnu\left( z p \right) u(0,z) dz. \nonumber
\end{align}
The solution of this problem reads
\begin{equation}
\label{barU_explicit_StripDom}
\baru = e^{-p^2\tau / 2} \left[ \bar{u}(0,p) + \frac{1}{2} \int_0^\tau e^{p^2s /2} \Psi(s) y^{\nu+ 1}(s) \Jnu(y(s) p) ds \right].
\end{equation}
By analogy with \citep{CarrItkin2020}, we can obtain the Fredholm equation of the first type for the function $\Psi(\tau)$. For doing so, let us set $p = i\lambda, \ \lambda \in \mathbb{R}$, tend $\tau$ to infinity and apply the formula $\Jnu(\iu x) = e^{\iu \nu \pi /2} I_{\nu}(x)$ which connects the Bessel function $J_\nu(x)$ with the modified Bessel function $I_\nu(x)$. This yields
\begin{equation} \label{Fred}
\int_0^\infty e^{-\lambda^2 s /2} \Psi(s) y^{\nu +1}(s) I_{\nu} \left(y(s) \lambda \right) ds  = - 2\int_0^{y(0)} q^{\nu + 1} I_{\nu}\left( q \lambda \right) u(0,q) dq.
\end{equation}
The solution of this integral equation $\Psi(\tau)$ can be found numerically on a grid by solving a system of linear equations, see e.g., \citep{CarrItkin2020} for the discussion on this subject and a numerical example. Once the function $\Psi(\tau)$ is found, it has to be substituted into \eqref{baruCauchy} to obtain the generalized transform of $u(\tau,z)$ in the explicit form. Then, if this transform can be inverted back, we solved the problem of pricing Up-and-Out barrier Call options for the CEV model with $\beta < 0$.

\subsubsection{The inverse transform}

As it was already mentioned, it is reasonable to seek the solution of the CEV problem in the class of the Bessel functions. Therefore, we represent the solution in the form
\begin{equation} \label{invTrStrip}
u(\tau, z) = z^{-\nu} \sum_{n = 1}^{\infty} \alpha_n(\tau) J_{|\nu|} \left( \frac{\mu_n z}{y(\tau)}\right)
\end{equation}
Here $\mu_n$ is an ordered sequence of the positive zeros of the Bessel function $\Jnu(\mu)$:
\[
J_{|\nu|}(\mu_n) = J_{|\nu|}(\mu_m) = 0, \quad \mu_n > \mu_m > 0, \quad n >m.
\]
Note, that the definition in \eqref{invTrStrip} automatically respects the vanishing boundary conditions for $u(\tau, z)$. We assume that this series converges absolutely and uniformly $\forall z \in  [0,y(\tau)]$ for any $\tau > 0$.

Applying the direct integral transform in \eqref{GITdef} to both parts of \eqref{invTrStrip}, and using a change of variables $z \to \hz = z y(\tau)$ yields
\begin{equation}
\frac{\bar{u}(\tau, p)}{y^2(\tau)} =  \sum_{n = 1}^{\infty} \alpha_n(\tau) \int_0^{1} \hz   J_{|\nu|} \left( \mu_n \hz\right)
\Jnu(p y(\tau) \hz) d\hz.
\end{equation}
The set of functions $\Jnu(\alpha \hz)$ with $\alpha \in \mu_n, \ n=1,\ldots,$ forms an orthogonal basis in the space $C[0,1]$ with the scalar product
\begin{equation} \label{scal}
\langle \Jnu(\alpha z), \Jnu(\beta z) \rangle =  2\int_0^1 \frac{z \Jnu(\alpha z) \Jnu(\beta z) dz}{J_{|\nu| + 1}(\alpha) J_{|\nu| + 1}(\beta)}  =
 {\left\{ \begin{array}{l}  1, \quad \alpha = \beta, \\ 0, \quad \alpha \neq \beta    \end{array}  \right.}
\end{equation}
Therefore, the explicit formula for each coefficient $\alpha_n(\tau)$ is straightforward
\begin{equation}
\alpha_n(\tau) = 2 \frac{\bar{u}\left(\tau, \mu_n /y(\tau)\right)}{y^2(\tau) J_{|\nu| + 1}^2 (\mu_n)}.
\end{equation}
Thus, the final solution for $u(\tau, z)$ reads
\begin{align} \label{Uexplicit}
u(\tau, z) &= 2 \frac{z^{-\nu}}{y^2(\tau)} \sum_{n = 1}^\infty \bigg[ \int_0^{y(0)} u(0,s) s^{\nu + 1}
 e^{-\frac{\mu_n^2}{2 y^2(\tau)}\tau} \frac{\Jnu ( \mu_n s  / y(\tau)) \Jnu(\mu_n z  / y(\tau))}{J_{|\nu| + 1}^2 (\mu_n)} ds \\
&+ \frac{1}{2}\int_0^\tau  y^{\nu+1}(s) \Psi(s)  e^{-\frac{\mu_n^2}{2 y^2(\tau)}(\tau - s)} \frac{\Jnu(\mu_n y(s) / y(\tau)) \Jnu(\mu_n z / y(\tau))}{J_{|\nu| + 1}^2 (\mu_n)} ds \bigg]. \nonumber
\end{align}
This expression can be also re-written in the form
\begin{align} \label{Uexplicit1}
u(\tau, z) &= 2 \frac{z^{-2\nu}}{y^2(\tau)}  \Big[ \int_0^{y(0)} s\ u(0,s)  \Theta_{|\nu|}\left(\frac{\sqrt{\tau}}{y(\tau)}, \frac{s}{y(\tau)}, \frac{z}{y(\tau)} \right)  ds \\
&+ \frac{1}{2}\int_0^\tau  y(\tau)  \Psi(s)  \Theta_{|\nu|}\left(\frac{\sqrt{\tau-s}}{y(\tau)}, \frac{y(s)}{y(\tau)}, \frac{z}{y(\tau)} \right) ds \Big], \nonumber
\end{align}
\noindent where we introduced a new function
\begin{equation} \label{thetaBess}
\Theta_{|\nu|}(\theta, x_1, x_2)   = \sum_{n = 1}^\infty e^{-\frac{\mu_n^2 \theta^2}{2}} (x_1 x_2)^\nu \frac{\Jnu ( \mu_n x_1) \Jnu(\mu_n x_2)}{J_{|\nu| + 1}^2 (\mu_n)}.
\end{equation}
The function $\Theta_{|\nu|}(\theta, x_1, x_2) $ is an analog of the Jacobi theta function which is a periodic solution of the heat equation. Indeed, in \citep{CarrItkin2020} the solution of a similar problem for the time-dependent OU model (so $\beta = -1,  \nu = -1/2$ and $|\nu| = 1/2$) with moving boundaries but for the heat equation has been obtained in terms of the theta functions. It can be checked that, if $\nu = 1/2$, we have
\begin{align}
\Theta_{1/2}\left(\frac{\sqrt{\tau}}{y(\tau)}, \frac{s}{y(\tau)}, \frac{z}{y(\tau)} \right) =
\frac{y(\tau)}{ 2} \left[ \theta_3\left(e^{-\frac{\pi^2 \theta^2}{8}}, \frac{n \pi(s-z)}{4 y(\tau)}\right) -
\theta_3\left(e^{-\frac{\pi^2 \theta^2}{8}}, \frac{n \pi(s+z)}{4 y(\tau)}\right) \right],
\end{align}
\noindent where $\theta_3(\omega, x)$ is the Jacobi theta function, \citep{mumford1983tata}. Accordingly, function $\Theta_{|\nu|}(\theta, x_1, x_2)$ is a periodic solution of the Bessel equation.

Alternatively to the Fredholm equation of the first kind in \eqref{Fred} which is ill-posed and requires special methods to solve it, see \citep{CarrItkin2020}, we can use a trick proposed in \citep{ItkinMuravey2020} and instead derive the Volterra equation of the second kind for the function $\Psi(\tau)$. For doing that, one needs to differentiate \eqref{Uexplicit} on $z$, and then let $z = y(\tau)$. This yields
\begin{align} \label{Volt2}
\Psi(\tau) &= -\frac{2}{y^{3+\nu}(\tau)} \sum_{n = 1}^\infty  (\mu_n + \nu) \bigg[ \int_0^{y(0)} u(0,s) s^{\nu+1}
 e^{-\frac{\mu_n^2 \tau}{2 y^2(\tau)}} \frac{\Jnu ( \mu_n s  / y(\tau))}{J_{|\nu| + 1} (\mu_n)} ds \\
&+ \frac{1}{2}\int_0^\tau y^{\nu + 1}(s)\Psi(s)   e^{-\frac{\mu_n^2 (\tau - s)}{2 y^2(\tau)}}\frac{\Jnu(\mu_n y(s) / y(\tau))}{J_{|\nu| + 1}(\mu_n)} ds \bigg]. \nonumber
\end{align}
This equation has to be solved numerically, again see \citep{ItkinMuravey2020} for the discussion and examples.

\subsubsection{Some approximations}

In some cases, \eqref{Volt2} can be solved asymptotically. For instance, one can apply the following approximations
\begin{align} \label{appr}
J_{|\nu|}^2(z) &+ J_{|\nu| + 1}^2(z) \approx \frac{2}{\pi z}, \qquad x \gg \nu,\quad
J_{|\nu| + 1}^2(\mu_n) \approx \frac{2}{\pi \mu_n}, \quad \mu_n \gg \nu \\
\Jnu(z) &\sim \sqrt{\frac{2}{\pi z}} \cos \left(z - \frac{2|\nu| +1}{4}\pi\right),\quad z \rightarrow \infty
\qquad \mu_n \approx \pi \left( n + \frac{2|\nu| + 1}{4}\right), \quad n \rightarrow \infty. \nonumber
\end{align}
Then the infinite sum in \eqref{Uexplicit} can be truncated up to keep first $N$ terms. The reminder (the error of this method) reads
\begin{align} \label{U_rem}
R(N, \tau, z) &= \frac{1}{z^{\nu + 1/2}y(\tau)}
\sum_{n = N + 1}^\infty \bigg\{ \\
& \int_0^{y(0)} s^{\nu + 1/2} u(0,s)  e^{-\frac{\pi^2 \lambda_n^2}{2 y^2(\tau)}\tau}
\left[\cos\left(\frac{\pi \lambda_n (s + z)}{y(\tau)} - 2\pi \lambda_n\right) +
\cos\left(\frac{\pi \lambda_n(s - z)}{y(\tau)} \right) \right] ds \nonumber \\
&+ \frac{1}{2}\int_0^\tau  y^{\nu+1/2}(s) \Psi(s)  e^{-\frac{\pi^2 \lambda_n^2}{2 y^2(\tau)}(\tau - s)}
\left[ \cos\left(\frac{\pi \lambda_n(y(s) + z)}{y(\tau)} - 2\pi \lambda_n\right) +
\cos\left(\frac{\pi \lambda_n(y(s) - z)}{y(\tau)} \right) \right] ds \bigg\}. \nonumber
\end{align}
Here $\lambda_n = n + (2\nu - 1) / 4$

A simple assessment shows that, since $-1 < \beta < 0$, from \eqref{PDE1} at typical values of the model parameters  we have $f(t) \approx f_1 \ll 1$. Assume that $H = const$. Hence, in \eqref{tr2}
\[
\phi = \frac{\log (2 \beta f_1 \tau +1)}{2 \beta  f_1}, \qquad y(\tau) = -\frac{H^{-\beta } \sqrt{2 \beta f_1 \tau +1}}{\beta }.
\]
Then it can be checked that the expression
\[ -\frac{\mu_n^2 \tau}{2 y^2(\tau)} \]
\noindent rapidly drops down for $-1 < \beta < 0, \ \tau > 0$, unless $\tau \ll 1$ and $H \gg 1$. Therefore, in this case just first few terms in the sum in \eqref{Volt2} would be a good approximation.

Another approximation can be proposed to compute function $\Theta_{|\nu|}(\theta, x_1, x_2)$ defined in \eqref{thetaBess}. The idea is that, as mentioned in above if $\tau \ll 1$ and $H = O(1)$, the first argument $\theta$ of this function is small, $\theta \ll 1$. Then
the function $c_0 = e^{-\frac{\mu_n^2 \theta^2}{2}}$ is small at large $n$, and $\frac{\mu_n^2 \theta^2}{2}$ is small at small $n$. Therefore,
for small $n$ we represent $c_0$ by using the \pade approximation $(k,1)$, and for large $n$ we replace the small values of $c_0$ with the same \pade approximation. To estimate how accurate is this trick, let us pick, for instance, $k=2$, so
\begin{equation} \label{pade}
c_0 \approx \frac{1}{1 + \frac{x}{2}} \left(1 - \frac{ x}{2}\right) + O(x^2), \qquad x = \frac{\mu_n^2 \theta^2}{2}.
\end{equation}
Thus, expanding the parenthesis into two terms, the function $(x_1 x_2)^{-\nu} \Theta_{|\nu|}(\theta, x_1, x_2)$ can be represented as
\begin{equation} \label{totApp}
(x_1 x_2)^{-\nu} \Theta_{|\nu|}(\theta, x_1, x_2) = A_1(\theta, x_1, x_2) + A_2(\theta, x_1, x_2).
\end{equation}
Now observe that, \citep{bateman1953higher}
\begin{align} \label{thetaApp}
A_1(\theta, x_1, x_2) &= \sum_{n = 1}^\infty \frac{\Jnu ( \mu_n x_1) \Jnu(\mu_n x_2)}{J_{|\nu| + 1}^2 (\mu_n)} \frac{1}{1 + \frac{\mu_n^2 \theta^2}{4}} = - \frac{4}{\theta^2}\sum_{n = 1}^\infty \frac{\Jnu ( \mu_n x_1) \Jnu(\mu_n x_2)}{J_{|\nu| + 1}^2 (\mu_n) (\gamma^2 - \mu_n^2)} \\
&=  \frac{\pi \Jnu(x_2 \gamma)}{\theta^2 \Jnu(\gamma)}\left[\Jnu(x_1 \gamma) Y_{|\nu|}(\gamma) - \Jnu(\gamma) Y_{|\nu|}(x_1 \gamma) \right] \nonumber \\
&=  \frac{2 \Inu(x_2 \barg)}{\theta^2 \Inu(\barg)}\left[\Knu(x_1 \barg) \Inu(\barg) - \Knu(\barg) \Inu(x_1 \barg) \right], \qquad \gamma = 2 \iu/\theta, \quad \barg = 2/\theta. \nonumber
\end{align}
Here we again used the formulae
\begin{equation*}
J_\nu(ix) = e^{\iu \nu \pi / 2} I_\nu(x), \qquad Y_\nu(ix) = e^{(\nu + 1) \pi  \iu / 2} I_\nu(x) - 2/\pi e^{-\nu \pi \iu / 2} K_\nu(x),
\end{equation*}
\noindent which connect the Bessel functions $J_{\nu}(x)$ and $Y_{\nu}(x)$ with the modified Bessel functions $I_\nu(x)$ and $K_\nu(x)$.

To compute the next term in \eqref{pade}
\begin{align} \label{thetaApp1}
A_2(\theta, x_1, x_2) &= -\frac{\theta^2}{4} \sum_{n = 1}^\infty \frac{\Jnu ( \mu_n x_1) \Jnu(\mu_n x_2)}{J_{|\nu| + 1}^2 (\mu_n)} \frac{\mu_n^2}{1 + \frac{\mu_n^2 \theta^2}{4}}
\end{align}
\noindent let us differentiate the LHS of \eqref{thetaApp} by $\theta$, so
\begin{align} \label{thetaApp2}
\fp{}{\theta} \sum_{n = 1}^\infty & \frac{\Jnu ( \mu_n x_1) \Jnu(\mu_n x_2)}{J_{|\nu| + 1}^2 (\mu_n)} \frac{1}{1 + \frac{\mu_n^2
\theta^2}{4}} = - \frac{\theta}{2}\sum_{n = 1}^\infty \frac{\Jnu ( \mu_n x_1) \Jnu(\mu_n x_2)}{J_{|\nu| + 1}^2 (\mu_n)} \frac{\mu_n^2}{\left(1 + \frac{\mu_n^2 \theta^2}{4}\right)^2} \\
&\approx - \frac{\theta}{2}\sum_{n = 1}^\infty \frac{\Jnu ( \mu_n x_1) \Jnu(\mu_n x_2)}{J_{|\nu| + 1}^2 (\mu_n)} \frac{\mu_n^2}{1 + \frac{\mu_n^2 \theta^2}{2}}. \nonumber
\end{align}

Therefore, the second term in \eqref{pade} takes the form
\begin{align}
A_2(\theta, x_1, x_2) &= - {\frac{\barg}{4}} \fp{}{\barg} \left\{
\frac{\barg^2 \Inu(x_2 \barg)}{\Inu(\barg)}\left[\Knu(x_1 \barg) \Inu(\barg) - \Knu(\barg) \Inu(x_1 \barg) \right] \right\},  \quad
\barg = 2\sqrt{2}/\theta,
\\
\fp{\Inu(\gamma)}{\gamma} &= \frac{1}{2} \left(I_{|\nu|-1} (\gamma) + I_{|\nu|+1} (\gamma) \right), \qquad
\fp{\Knu(\gamma)}{\gamma} = \frac{1}{2} \left(K_{|\nu|-1} (\gamma) + K_{|\nu|+1} (\gamma) \right). \nonumber
\end{align}

A similar approximation can be developed for \eqref{Volt2} by using the identity, \citep{bateman1953higher}
\begin{equation} \label{thetaApp3}
\sum_{n = 1}^\infty \frac{\mu_n \Jnu( \mu_n x) }{(\mu_n^2 - k^2) J_{|\nu| + 1}(\mu_n)} = \frac{\Jnu(k x)}{2\Jnu( k)}.
\end{equation}

\begin{figure}[!htb]
\vspace{-0.1in}
\begin{center}
\includegraphics[totalheight=3.5in]{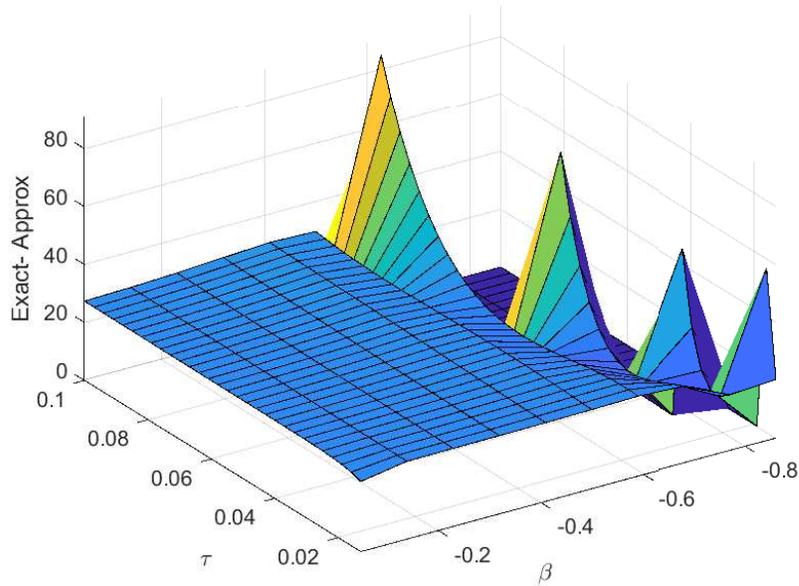}
\caption{Comparison of the exact and approximated values of the function $G_\nu(x_1,x_2)$.}
\label{Fig1}
\end{center}
\end{figure}

As an example, let consider the CEV model with constant parameters given in Table~\ref{Tab1}.
\begin{table}[!htb]
\begin{center}
\caption{Parameters of the test.}
\label{Tab1}
\begin{tabular}{|c|c|c|c|c|c|c|}
\hline
$r$ & $q$ & $\sigma$ & $H$ & $N$ & $x_1/y(\tau)$ & $x_2/y(\tau)$ \\
\hline
0.02 & 0.01 & 0.5 & 0.2 & 100 & 0.5 & 0.5 \\
\hline
\end{tabular}
\end{center}
\end{table}
In Fig.~\ref{Fig1} we present the difference of two values of the function $G_\nu(x_1,x_2) = (x_1 x_2)^{-\nu} \Theta_{|\nu|}(\theta, x_1, x_2)$: one computed by the definition in \eqref{thetaBess} using the first $N$ terms in the sum; and the other computed by using the approximation in \eqref{totApp}. It can be seen that the latter approximation provides an accuracy about 25\% except the area where both $\tau$ and $\beta$ are simultaneously kind of large, and hence, the value of $G_\nu(x_1,x_2)$ is very small.

This approach can be further developed by increasing $k$ in the \pade approximations $(k,1)$, and computing the consecutive terms in the expansion by using the same trick as in above. In other words, to get the next term in the numerator of the \pade approximation we can differentiate the LHS of \eqref{thetaApp} twice on $\theta$, and then express this derivative via the RHS of \eqref{thetaApp}, etc.

\subsubsection{Connection to the first passage time problem} \label{FPT}

Let us consider the following stopping moment
\[
T^{y}_{x_0} = \inf \left\{\tau \geq 0, X_\tau \geq y(\tau) \right\},
\]
\noindent where $X_\tau$ is the Bessel process defined by \eqref{BesProc} and originated from $X_0 = x_0$. From standard results in probability theory the p.d.f. $\rho_{x_0}^{y}(\tau)$ of the moment $T^{y}_{x_0}$ can be found via the Fokker--Planck--Kolmogorov equation associated with the process $X_\tau$:
\begin{align}
\label{eqFPKE}
    \frac{1}{2} \sop{F_{x_0}(\tau, x)}{x} - \frac{b}{x} \fp{F_{x_0}(\tau, x)}{x} + \frac{b}{x^2} F_{x_0}(\tau, x) = \fp{F_{x_0}(\tau, x)}{\tau}
    \\ \nonumber
    F_{x_0}(\tau, y(\tau)) = 0 , \quad F_{x_0}(\tau, 0) = \delta(x - x_0)
\end{align}
here $\delta(x- x_0)$ is the Dirac measure at the point $x_0$. The density $\rho_{x_0}^{y}(\tau)$ has the following representation:
\begin{equation}
  \rho_{x_0}^{y}(\tau) = \frac{1}{2} \fp{F_{x_0}(\tau,x)}{x} \bigg|_{x = y(\tau)}.
\end{equation}
The solution of the problem \eqref{Bess1} and the function $\Psi(\tau)$ can be represented in terms of the function $F$:
\begin{equation}
\label{eqFPT2Psi}
    u(z,\tau) = z^{-2b}\int_0^{y(0)} s^{2b} u(0,s) F_s(\tau, z) ds, \quad
    \Psi(\tau) = 2y(\tau)^{-2b} \int_0^{y(0)} s^{2b} u(0,s) \rho_{s}^{y}(\tau) ds
\end{equation}
For the boundary moving linearly in time $y(\tau) \approx \alpha + \beta \tau$ it is possible to propose  the following approximation for the function $\Psi(\tau)$
\begin{equation}
\Psi(\tau)  \approx
2y(\tau)^{-(2\nu +1)}
\int_0^{y(0)} s^{2\nu +1} u(s,0) \frac{e^{\frac{\beta}{2\alpha}\left(\alpha^2 - s^2\right) + \frac{b^2}{2} \tau}}
{\left(\alpha + \beta \tau \right)^{\nu+ 2}}
\sum_{n = 1}^{\infty}
\frac{s^{-\nu} \mu_n \Jnu(\mu_n z / \alpha)}
{\alpha^{-2\nu} J_{|\nu| + 1}(\mu_n} e^{-\mu_n^2 \frac{\tau}{2\alpha (\alpha + \beta \tau)}} ds.
\end{equation}
This is due to the fact that the p.d.f $\rho_{x_0}^{y}(\tau)$ of the first hitting time of the line $a\tau + b$ for the Bessel process is known explicitly, \citep{AliliPatie2010}.

Note, that a similar problem for the OU process with both constant and time-dependent coefficients has been studied in \citep{LiptonKau2020-2}.
The authors considered the first hitting time density to a moving boundary for a diffusion process, which satisfies the Cherkasov condition, and hence, can be reduced to a standard Wiener process. They give two complementary (forward and backward) formulations of this problem and provide semi-analytical solutions for both by using the method of heat potentials.

\subsection{Domain $z > y(\tau)$.} \label{fixDomGITInf}

To recall, this problem occurs in both the CEV model with $0 < \beta < 1$ and the CIR model. We will construct the solution of this problem by using the Weber--Orr transform
\begin{align} \label{WeberOrrTransform_def}
\baru(\tau,p) &= \int_{y(\tau)}^{\infty} z^{\nu + 1} W(\tau, p,z) u(\tau, z) dz \\
u(\tau, z) &= z^{-\nu} \int_{0}^{\infty} \frac{ p W(\tau, p, z)}{V(\tau, p)} \baru(\tau,p)  dp. \nonumber
\end{align}
The kernel $W(a, b)$ and the function $V(p)$ are defined as follows, \citep{bateman1953higher}
\begin{align} \label{WeberOrrKernel_def}
W(\tau, a, b) &= \Jnu(a b) \Ynu (a y(\tau)) -  \Ynu(a b) \Jnu(a y(\tau)), \\
V(\tau, p) &= \Jnu^2(p y(\tau)) + \Ynu^2(p y(\tau)). \nonumber
\end{align}
The definitions in \eqref{WeberOrrKernel_def} are generalizations of the Pythagorean and Angle sum identities for trigonometric functions to the case of cylinder functions $\Jnu$ and $\Ynu$.  The functions $W(\tau, a, b)$ as the functions of the second argument $a$ also form an orthogonal basis in the space $C[y(\tau), \infty)$ for all $\tau > 0$.

However, we cannot apply this transform directly to the Bessel equation due to fact that the kernel is time-dependent. Therefore, we propose to represent $\bar{u}$ as the weighted sum of the following transforms
\begin{equation}
 \baru_J(p,\tau) = \int_{y(\tau)}^{\infty} z^{\nu + 1} \Jnu(z p) u(\tau, z) dz,
\qquad
\baru_Y(p,\tau) = \int_{y(\tau)}^{\infty} z^{\nu + 1} \Ynu(z p) u(\tau, z) dz,
\end{equation}
\noindent so
\begin{equation}
\baru\tau, p) = \baru_J(\tau,p) \Ynu(y(\tau) p) - \baru_Y(\tau,p) \Jnu(y(\tau) p)
\end{equation}
The explicit formulae for $\baru_J$ and $\baru_Y$ read
\begin{align}
\baru_J(\tau,p) =e^{-p^2\tau / 2} \bigg[\baru_J(0,p) +  \frac{1}{2}\int_0^\tau e^{p^2s /2} \Psi(s) y^{\nu + 1}(s) \Jnu \left(y(s) p \right) ds \bigg], \\
\baru_Y(\tau,p) =e^{-p^2\tau / 2} \bigg[\baru_Y(0,p) +  \frac{1}{2}\int_0^\tau e^{p^2s /2} \Psi(s) y^{\nu + 1}(s) \Ynu \left(y(s) p \right) ds \bigg], \nonumber
\end{align}
\noindent where $\Psi(\tau)$ is defined in \eqref{sumJ}. Using the inversion formula from \eqref{WeberOrrTransform_def} we immediately get the explicit formula for $u(\tau, z):$
\begin{align} \label{uWebOrrTrs}
u(z,\tau) &= z^{-\nu} \int_0^\infty \int_{y(0)}^\infty s^{\nu + 1}u(0, s) e^{-\frac{p^2 \tau }{2} } \frac{W(\tau,p, z) W(\tau,p, s)}{V(\tau,p)} p\, dp\, ds \\
&+ \frac{z^{-\nu}}{2}\int_0^\infty \int_{0}^\tau y^{\nu + 1}(s) \Psi(s) e^{-\frac{p^2}{2}(\tau - s)}
\frac{W(\tau,p, z) W(\tau,p, y(s))}{V(\tau,p)} p\, dp\, ds. \nonumber
\end{align}
By analogy with the previous section, the function $\Psi(\tau)$ solves the Fredholm equation of the first kind
\begin{equation} \label{Fred2}
\int_0^\infty e^{-\lambda^2 s /2} \Psi(s) y^{\nu +1}(s) K_{\nu} \left(y(s) \lambda \right) ds  = - 2\int_{y(0)}^{\infty} q^{\nu + 1} K_{\nu}\left( q \lambda \right) u(0,q) dq,
\end{equation}
\noindent or the Volterra equation of the second kind
\begin{align}
\label{VolterraGIT}
\Psi(\tau) &= y^{-\nu}(\tau) \bigg [\int_0^\infty \int_{y(0)}^\infty q^{\nu + 1}u(0, q)e^{-\frac{p^2 \tau }{2} }
\frac{Q(\tau,p)  W(\tau, p, q)}{V(\tau, p)} dp\, dq \bigg] \\
&+ \frac{1}{2}\int_0^\infty \int_{0}^\tau y^{\nu + 1}(s) \Psi(s) e^{-\frac{p^2}{2}(\tau - s)}
	\frac{Q(\tau,p) W(\tau, p, y(s))}{V(\tau, p)}  dp\, ds. \nonumber
\end{align}
Here
\begin{equation}
Q(\tau,p) = J_{\nu+ 1} (py(\tau)) Y_{\nu} (py(\tau)) - Y_{\nu+ 1} (py(\tau)) J_{\nu} (py(\tau)).
\end{equation}

In some cases, \eqref{uWebOrrTrs} can be solved asymptotically. For instance, one can apply the following approximations
\begin{align} \label{appr2}
\Jnu^2(z) + \Ynu^2(z) &\approx
\frac{2}{\pi z} \sum_{k = 0} ^\infty \frac{(2 k - 1)!!}{2^k z^{2k}} \frac{\Gamma(|\nu| + k + 1/2)}{k! \Gamma(|\nu| - k + 1/2)}, \quad
\Jnu^2(z) + \Ynu^2(z) \sim \frac{2}{\pi z}, \quad z \rightarrow \infty \\
\Ynu(z) &\sim \sqrt{\frac{2}{\pi z}} \sin \left(z - \frac{2|\nu| + 1}{4}\right), \quad z \rightarrow \infty,
\quad W(\tau, a,b) \sim \frac{2 \sin \left(a [b - y(\tau)] \right)}{\pi a \sqrt{b y(\tau)}}, \quad a \rightarrow \infty. \nonumber
\end{align}
Then, the outer integral in \eqref{uWebOrrTrs} can be truncated from above and approximated by the integral over the domain $[0, P], \ 0 < P < \infty$. The reminder (the error of this method) reads
\begin{align}
R(P,\tau, z) &= z^{-\nu-1/2} \int_P^\infty  \bigg \{ \int_{y(0)}^\infty s^{\nu + 1/2} u(0, s)e^{-\frac{p^2 \tau }{2}}
\sin\left(p[z -y(\tau)]\right)  \sin\left(p[s -y(\tau)]\right) ds \\
&+\frac{1}{2} \int_{0}^\tau y^{\nu + 1/2}(s) \Psi(s) e^{-\frac{p^2}{2}(\tau - s) }
\sin\left(p[z -y(\tau)]\right)  \sin\left(p[y(s) -y(\tau)]\right) ds \bigg\} dp. \nonumber
\end{align}
Introducing the new function $\Upsilon(P, t, \eta)$
\[ \Upsilon(P, t, \eta) = e^{-\frac{\eta^2}{2t}} \left[\erfc\left(\frac{Pt + \iu \eta}{\sqrt{2 t }}\right) + \erfc\left(\frac{Pt - \iu \eta}{\sqrt{2 t }}\right)\right],
\]
\noindent and taking into account the identity
\[ \int_P^\infty e^{-\frac{p^2 t}{2}} \cos(\eta p)\, dp  =   \frac{1}{2} \sqrt{\frac{\pi}{2 t}}\Upsilon(P, t, \eta),
\]
\noindent we obtain the following explicit representation for $R(P, \tau, z)$
\begin{align}
R(P,\tau, z) &=
\frac{z^{-\nu-1/2} \sqrt{\pi}}{4\sqrt{2}} \bigg \{
\int_{y(0)}^\infty \frac{s^{\nu + 1/2} u(0,s)}{\sqrt{\tau}} \left[ \Upsilon(P, \tau, z-s) - \Upsilon(P, \tau, z+s - 2y(\tau))
\right] ds \\
&+ \frac{1}{2} \int_{0}^{\tau} \frac{y^{\nu + 1/2}(s) \Psi(s)}{\sqrt{\tau - s}}
\left[ \Upsilon(P, \tau - s, z-y(s)) - \Upsilon(P, \tau-s, z+y(s) - 2y(\tau)) \right] ds \bigg\}. \nonumber
\end{align}
Now we show that under the assumptions
\begin{equation}
\int_{y(0)}^\infty q^{\nu + 1/2} u(0,q) dq \leq  M_1, \quad \int_{y(0)}^\infty q^{2\nu + 1} u(0,q) dq \leq M_2, \quad M_1, M_2 - const,
\end{equation}
\noindent we can set the upper limit of integration $P$ such that $|R(P,\tau, z)| < \epsilon$ for any  $\epsilon >0$. Indeed, using the following inequalities (see \eqref{eqFPT2Psi}):
\begin{equation}
\frac{1}{\sqrt{2t}} \Upsilon(P,t,\eta) \leq \erfc(P), \quad
\Psi(\tau) \leq 2 y(\tau)^{-(1+2\nu)} \int_{y(0)}^{\infty} q^{2\nu + 1} u(0, q) dq,
\end{equation}
\noindent yields
\begin{equation}
|R(P,\tau, z)| \leq \frac{z^{-\nu-1/2} \sqrt{\pi} \erfc(P) }{2}
\bigg \{
\int_{y(0)}^\infty q^{\nu + 1/2} u(0,q) dq +
\int_{0}^{\tau} y^{-\nu -1/2}(s)ds \int_{y(0)}^\infty q^{2\nu + 1} u(0,q) dq
\bigg \}.
\end{equation}
Since the second integral is bounded for any $\tau$, we obtain the following inequality:
\begin{equation}
|R(P,\tau, z)| \leq \frac{z^{-\nu-1/2} \sqrt{\pi} \erfc(P)}{2}
\left(M_1 + M_2 M_3(\tau) \right), \quad M_3(\tau) = \int_{0}^{\tau} y^{-\nu -1/2}(s)ds.
\end{equation}

\section{Numerical experiments} \label{numRes}

Similar to \citep{ItkinMuravey2020}, to check performance and accuracy of the proposed methods we construct the following test. We consider Up-and-Out Barrier Call option written on the underlying stock which follows the CEV process with $0 < \beta < 1$. This case is described at the end of Section~\ref{SecBO}. To recall, after the change of variables proposed in that Section is done, the problems is transformed to the solution of the Bessel PDE at the domain $\bar{z} \in [y(\tau), \infty)$ with the boundary and initial conditions given in \eqref{bc2}, \eqref{tc2}\footnote{Hence, in new variables the Up-and-Out option transforms to the Down-and-Out option.}

In this test we use the explicit form of parameters $r(t),q(t), \sigma(t)$
\begin{equation} \label{ex}
r(t) = r_0 - r_k (a+t), \qquad q(t) = q_0 - q_k(a+t), \qquad \sigma(t) = \sigma_0 \sqrt{a+t},
\end{equation}
\noindent where $r_0, q_0, \sigma_0, r_k, q_k, \sigma_k$ are constants. We also assume $r_0 = q_0$, and $H - const$. With these definitions one can find
\begin{align} \label{exCoef}
\phi(t) &= -\frac{1}{2} \sigma_0^2 (t-T) (2 a+t+T), \qquad \phi(\tau)  = \frac{\sigma_0^2}{2 \beta  (q_k - r_k)}
\log \left( \frac{\sigma_0^2 + 2 \beta  (q_k-r_k)\tau}{\sigma_0^2}\right), \\
F(\phi) &=\frac{\sqrt{2 \beta \tau (q_k-r_k) + \sigma_0^2}}{\sigma_0}, \qquad y(\tau) = F(\phi ) H^{-\beta }/\beta. \nonumber
\end{align}

We approach pricing the Up-and-Out barrier Call option in the CEV model twofold. First, as a benchmark we solve the PDE in \eqref{PDE}  by using a finite-difference (FD) scheme of the second order in space and time. We use the Crank-Nicolson scheme with few first Rannacher steps on a non-uniform grid compressed close to the barrier level, see \citep{ItkinBook}. Accordingly, our domain in $S$ space is $S \in [0,H]$
\footnote{A similar approach can be developed for the Down-and-Out options with $S \in [L,\infty)$. Then instead of truncating the infinite semi-interval, one can transform it to the fixed interval $[-1,1)$ or to $[0,1)$ and solve a modified PDE on the new interval. The boundary behavior of the solution can be obtained using Fichera theory and/or Greeen's integral formula, see \citep{Duffy2014} where this was done in many cases and proved that this approach works well.}.

Alternatively, we apply the method of Bessel potentials (BP) developed in Section~\ref{infDom} to solve the Bessel PDE \eqref{Bess}.
For doing so, first we solve the Volterra equation in \eqref{Abel2k} where the kernel is approximated on a rectangular grid $M \times M$, and
the integral is computed using the trapezoidal rule. This implies solving the following system of linear equations
\begin{equation} \label{matEq}
\| \varsigma \| = (I + P) \| \Psi \|.
\end{equation}
Here $\| \Psi \|$ is the vector of discrete values of $\Psi(\tau), \ \tau \in \left[0, \tau(t)\Big|_{t=0}\right]$ on a grid with $M$ nodes, $\| \varsigma \|$ is a similar vector of $\varsigma(\tau)$, $I$ is the unit $M\times M$ matrix, and $P$ is the $M\times M$ matrix  of the kernel values on the same grid. Note, that the matrix $P$ is lower triangular. Therefore, solution of \eqref{matEq} can be done with complexity $O(M^2)$.

As the kernel (and so the matrix $P$) doesn't depend of strikes $K$, but only the function $\varsigma(\tau)$, \eqref{matEq} can be solved simultaneously for all strikes by inverting the matrix $I + P$ with the complexity $O(M^2)$, and then multiplying it by vectors $\| \varsigma \|_k, \ k=1,\ldots,\bar{k}$,  $\bar{k}$ is the total number of strikes. Therefore, the total complexity of this step remains $O(\bar{k}M^2)$, but this operation, however, can be vectorized in $k$.  The Volterra equation could also been solved by iterative methods, but with almost the same complexity, see discussion in \citep{ItkinMuravey2020}.

\begin{table}[!htb]
\begin{center}
\caption{Parameters of the test.}
\label{tab1}
\begin{tabular}{|c|c|c|c|c|c|c|c|c|c|}
\hline
$r_0$ & $q_0$ & $\sigma_0$ & $r_k$ & $q_k$ & $\sigma_k$ & $a$ & $H$  & $S$\\
\hline
0.01 & 0.01 & 0.3 & 0.01 & 0.005 & 0.2 & 1.0 & 100 & 70  \\
\hline
\end{tabular}
\end{center}
\end{table}

The model parameters for this test parameters are presented in Table~\ref{tab1}. We run the test for a set of maturities $T \in [1/12, 0.3,0.5,1]$ and strikes $K \in [59, 64, 69, 74, 79, 84]$. The Up-and-Out barrier Call option prices computed in such an experiment are presented in Table~\ref{tab2}.

\subsection{Comparison with the BP method}

\begin{figure}[!htb]
\vspace{-0.1in}
\begin{center}
\caption{Up-and-Out barrier Call option price computed by using the BP method.}
\label{figPrice}
\includegraphics[totalheight=3.5in]{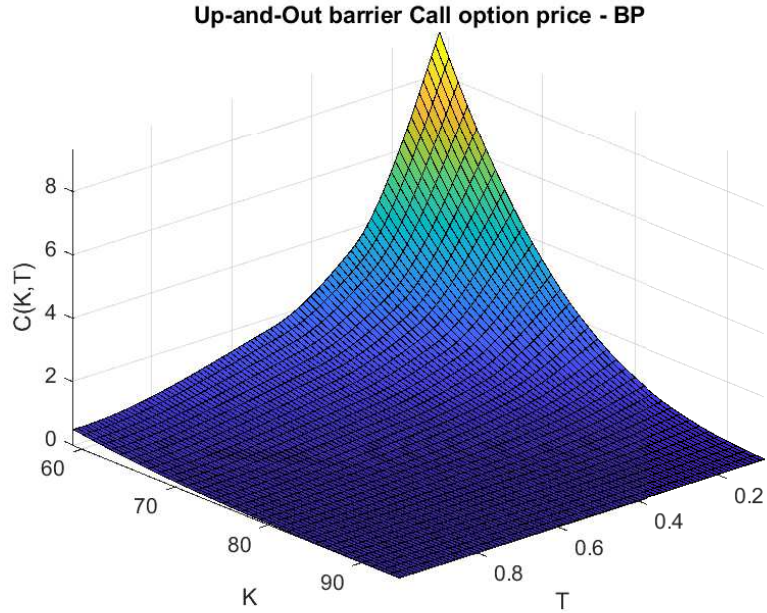}
\end{center}
\end{figure}

\begin{figure}[!htb]
\vspace{-0.1in}
\begin{center}
\caption{Percentage difference of Up-and-Out barrier Call option prices computed by using the BP and FD methods.}
\label{figDif}
\includegraphics[totalheight=3.5in]{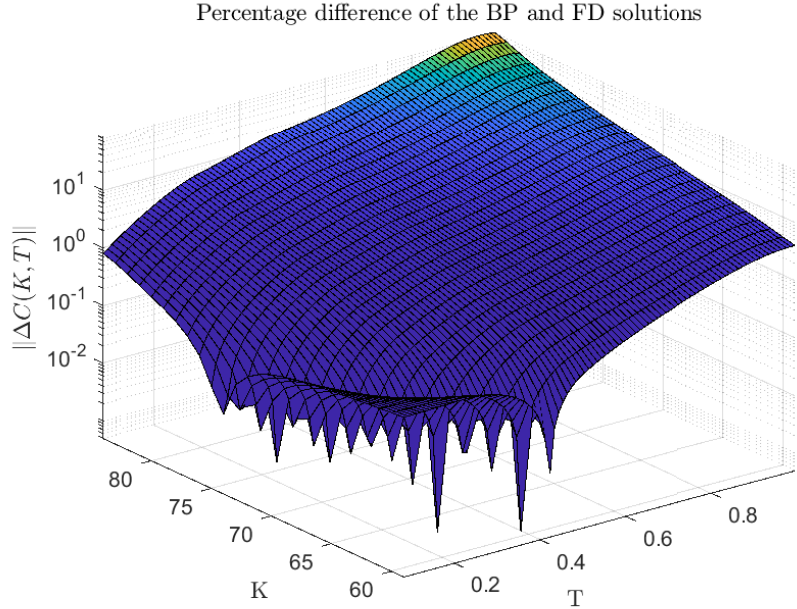}
\end{center}
\end{figure}

The same results computed by using the BP method are displayed in Fig.~\ref{figPrice} for the option prices. Also, in Fig.~\ref{figDif} the percentage difference between the prices obtained by using the BP and FD methods is presented as a function of the option strike $K$ and maturity $T$. Here to provide a comparable accuracy we run the FD solver with 101 nodes in space $S$ and 100 steps in time $t$. Otherwise the quality of the FD solution is not sufficient.

\begin{table}[!htb]
  \centering
  \caption{Up-and-Out barrier Call option prices computed by using the BP and FD methods.}
    \label{tab2}%
  \scalebox{0.8}{
  \renewcommand{\arraystretch}{1.3}
    \begin{tabular}{!{\vrule width 1pt}c!{\vrule width 1pt}r|r|r|r!{\vrule width 1pt}r|r|r|r!{\vrule width 1pt}r|r|r|r!{\vrule width 1pt}}
    \toprule
          & \multicolumn{4}{c!{\vrule width 1pt}}{BP} & \multicolumn{4}{c!{\vrule width 1pt}}{\textbf{FD}} & \multicolumn{4}{c!{\vrule width 1pt}}{\textbf{Difference \%}} \\
    \bottomrule
    \textbf{K\textbackslash{}T}
    & \multicolumn{1}{c|}{\bf{0.0833}} & \multicolumn{1}{c|}{\bf{0.3}} & \multicolumn{1}{c|}{\bf{0.5}} & \multicolumn{1}{c!{\vrule width 1pt}}{\bf{1.0}}
    & \multicolumn{1}{c|}{\bf{0.0833}} & \multicolumn{1}{c|}{\bf{0.3}} & \multicolumn{1}{c!{\vrule width 1pt}}{\bf{0.5}} & \multicolumn{1}{c!{\vrule width 1pt}}{\bf{1.0}}
    & \multicolumn{1}{c|}{\bf{0.0833}} & \multicolumn{1}{c|}{\bf{0.3}} & \multicolumn{1}{c|}{\bf{0.5}} & \multicolumn{1}{c!{\vrule width 1pt}}{\bf{1.0}} \\
    \bottomrule
    \bf{59}    & 9.3192 & 3.3642 & 1.6845 & 0.4976 & 9.2924 & 3.3554 & 1.6884 & 0.5175 & 0.2876 & 0.2604 & -0.2321 & -3.9899 \\
    \hline
    \bf{64}    & 6.2167 & 2.1795 & 1.0671 & 0.3038 & 6.2025 & 2.1831 & 1.0793 & 0.3252 & 0.2286 & -0.1654 & -1.1438 & -7.0291 \\
    \hline
    \bf{69}    & 3.8402 & 1.3219 & 0.6339 & 0.1731 & 3.8341 & 1.3319 & 0.6494 & 0.1931 & 0.1597 & -0.7624 & -2.4444 & -11.5443 \\
    \hline
    \bf{74}    & 2.1608 & 0.7350 & 0.3450 & 0.0891 & 2.1605 & 0.7477 & 0.3606 & 0.1061 & 0.0118 & -1.7293 & -4.5240 & -19.1029 \\
    \hline
    \bf{79}    & 1.0746 & 0.3612 & 0.1652 & 0.0388 & 1.0775 & 0.3736 & 0.1787 & 0.0522 & -0.2700 & -3.4102 & -8.1779 & -34.6241 \\
    \hline
    \bf{84}    & 0.4448 & 0.1462 & 0.0641 & 0.0121 & 0.4484 & 0.1561 & 0.0743 & 0.0216 & -0.7971 & -6.7649 & -15.9277 & -78.1360 \\
    \bottomrule
    \end{tabular}%
    }
\end{table}%

\leavevmode \\

It can be seen that the agreement of both methods is good (less than 1\%) if the option price is not too small which happens when the strike $K$ is close to the barrier or at high maturities. In this case, as this is seen from Table~\ref{tab2}, the relative difference becomes large, but the absolute difference of two methods is about one cent, which is almost insignificant. Obviously, such cases are a challenge for any FD method, as at $t=T$ there is a jump in the initial condition at the boundary, and the first derivative of the solution doesn't exists in this point.

As far as performance of both methods is concerned, to decrease the elapsed time for the FD method instead of \eqref{PDE} we solve the corresponding forward PDE. Therefore, prices of all options for a given set of strikes and maturities could be obtained within one sweep. This also requires $\bar{m}\times \bar{k}$ integrations of the  product of thus found density function with the payoff function, where $\bar{m} $ is the total number of maturities, and $\bar{k}$ is the total number of strikes. In this test the elapsed time for the FD method is, on average, 140 msec.

For the BP method, since the expression for $\varsigma(\tau)$ in \eqref{qP} is not known in closed form, we compute this integral numerically by using the Simpson quadratures. Nevertheless, to make the results accurate, we need to increase the number of the grid nodes $M$. As compared with \citep{ItkinMuravey2020}, where a similar expression for the Hull-White model could be computed in closed form, and $M=20$ provided a sufficient accuracy, here we need to take $M=100$.  Nevertheless, the elapsed time, on average, is 70 msec, i.e. twice faster than the FD method for the forward equation.

Decreasing $M$ almost doesn't impact the accuracy of the method at large values of the options prices, while slightly deteriorates the quality of the feed at large maturities and large strikes. Changing $M$ from 100 to 70 drops down the elapsed time to 50 msec. However, for the FD scheme decreasing the grid to $50\times50$ drops down the elapsed time to 30 msec while almost twice increasing the error for small maturities.
Overall, we can conclude that the method of BP demonstrates, at least, same performance as the forward FD solver.

\subsection{Comparison with the method of general integral transform (GIT)}

Here we solve the same problem by using the GIT method developed in Section~\ref{infDom}. Our numerical scheme is similar to that for the BP method: first we solve the Volterra equation \eqref{VolterraGIT} and then compute the value of the integrals in \eqref{uWebOrrTrs} using a trapezoidal rule. The inner integrals in the first summand in \eqref{VolterraGIT} and \eqref{uWebOrrTrs} can be computed explicitly for the payoff \eqref{tc1} via the following formulas \citep{GR2007}
\begin{alignat*}{2}
\int_0^1 x^{\nu + 1} & J_{\nu} (ax) dx = a^{-1} J_{\nu+ 1} (a),  &&\quad \Re(\nu) > -1, \\
\int_0^1 x^{1 -\nu}   & J_{\nu} (ax) dx = \frac{a^{\nu- 2}}{2^{\nu - 1} \Gamma(\nu)} - a^{-1}J_{\nu - 1}(a), &&\quad \Re(\nu) < 1, \\
\int_0^1 x^{\nu + 1} & Y_{\nu} (ax) dx = a^{-1} Y_{\nu+ 1} (a) + 2^{\nu+ 1} a^{-\nu- 2} \Gamma(\nu + 1), &&\quad \Re(\nu) > -1, \\
\int_0^1 x^{1 -\nu}   & Y_{\nu} (ax) dx = \frac{a^{\nu- 2}\cot(\nu \pi)}{2^{\nu - 1} \Gamma(\nu)} - a^{-1}Y_{\nu - 1}(a), &&\quad \Re(\nu) < 1.
\end{alignat*}
However, for the second summands we have to numerically compute two-dimensional integrals containing a lot of special functions.

We run the same test described in above, and the results of this numerical experiment are presented in Table~\ref{tab3} and also in Fig.~\ref{figDif} which depicts the percentage difference between the prices obtained by using the GIT and FD methods. For this test we use $M = 10$ steps in time. This algorithm was implemented in python.

\begin{table}[htb]
  \centering
  \caption{Up-and-Out barrier Call option prices computed by using the GIT and FD methods.}
    \label{tab3}%
  \scalebox{0.8}{
  \renewcommand{\arraystretch}{1.3}
    \begin{tabular}{!{\vrule width 1pt}c!{\vrule width 1pt}r|r|r|r!{\vrule width 1pt}r|r|r|r!{\vrule width 1pt}r|r|r|r!{\vrule width 1pt}}
    \toprule
          & \multicolumn{4}{c!{\vrule width 1pt}}{GIT} & \multicolumn{4}{c!{\vrule width 1pt}}{\textbf{FD}} & \multicolumn{4}{c!{\vrule width 1pt}}{\textbf{Difference \%}} \\
    \bottomrule
    \textbf{K\textbackslash{}T}
    & \multicolumn{1}{c|}{\bf{0.0833}} & \multicolumn{1}{c|}{\bf{0.3}} & \multicolumn{1}{c|}{\bf{0.5}} & \multicolumn{1}{c!{\vrule width 1pt}}{\bf{1.0}}
    & \multicolumn{1}{c|}{\bf{0.0833}} & \multicolumn{1}{c|}{\bf{0.3}} & \multicolumn{1}{c!{\vrule width 1pt}}{\bf{0.5}} & \multicolumn{1}{c!{\vrule width 1pt}}{\bf{1.0}}
    & \multicolumn{1}{c|}{\bf{0.0833}} & \multicolumn{1}{c|}{\bf{0.3}} & \multicolumn{1}{c|}{\bf{0.5}} & \multicolumn{1}{c!{\vrule width 1pt}}{\bf{1.0}} \\
    \bottomrule
    \bf{59}  & 9.3617 &	3.3532 & 1.6865 & 0.5158
             & 9.2924 & 3.3554  & 1.6884  & 0.5175
             & 0.71108 & -0.2191 & -0.3634 & -0.7888
             \\
    \hline
    \bf{64}    & 6.2737 & 2.1817 & 1.0781 & 0.3241
               & 6.2025 & 2.1831  & 1.0793  & 0.3252
               & 1.1101 & -0.2149 & -0.3654 & -0.7997 \\
    \hline
    \bf{69}    & 3.90483 & 1.3315 & 0.6488 & 0.1924
               & 3.8341 & 1.3319 & 0.6494   & 0.1931
               & 1.7729	& -0.20803 & -0.3556 & -0.8132 \\
    \hline
    \bf{74}    & 2.2194 & 0.7475 & 0.3604 & 0.1057
               & 2.1605 & 0.7477 & 0.3606 & 0.1061
               & 2.6313 & -0.1879 & -0.3318 & -0.8992 \\
    \hline
    \bf{79}    & 1.1171 & 0.3736 & 0.1786 & 0.0519
               & 1.0775 & 0.3736 & 0.1787 & 0.0522
               & 3.5572 & -0.1327 & -0.3129 & -1.0134 \\
    \hline
    \bf{84}    & 0.4692 & 0.1563 & 0.0742 & 0.0214
               & 0.4484 & 0.1561 & 0.0743 & 0.0216
               & 4.4718 & 0.0044 & -0.2848 & -1.1541 \\
    \bottomrule
    \end{tabular}%
    }
\end{table}%

\begin{figure}[!htb]
\vspace{-0.1in}
\begin{center}
\caption{Percentage difference of Up-and-Out barrier Call option prices computed by using the FD and GIT methods.}
\label{figDifGIT}
\includegraphics[totalheight=3.5in]{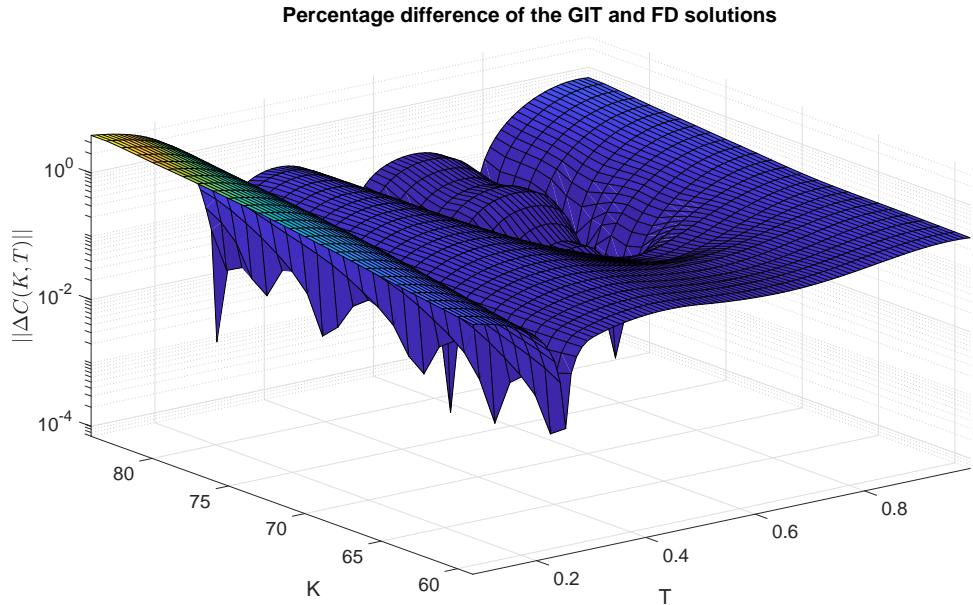}
\end{center}
\end{figure}

It can be seen that this method produces very accurate results at high strikes and maturities (i.e. where the option price is relatively small) in contrast to the BP method. This can be verified by looking at the exponents in \eqref{uWebOrrTrs} which are proportional to the time $\tau$. Contrary, when the price is higher (short maturities, low strikes) the GIT method is slightly less accurate than the BP method, as in \eqref{Abel2k} the exponent is inversely proportional to $\tau$. Obviously, the accuracy of the GIT method increases when $M$ increases.

This situation is well investigated for the heat equation with constant coefficients. As applied to pricing double barrier options, it is described in \citep{Lipton2002}. There exist two representation of the solution: one - obtained by using the method of images, and the other one - by the Fourier series. Despite both solutions are equal as the infinite series, their convergence properties are different. In particular, the Fourier method is superior when the difference between the upper $H$ and lower $L$ barriers is small. and the time is relatively large. And the image expansion should be used otherwise.

In this paper we come to a similar principle for the time-dependent problems, and not just for the heat equation but also for the Bessel one. Thus, it is important  that both the BP and GIT methods don't duplicate but rather compliment each other.

The speed of our python implementation is a bit slower than that for BP (approx. 0.158 sec). However, the latter method was implemented in Matlab. It is known that linear algebra in python (numpy) is almost  3 times slower than that in Matlab. Therefore, performance of both the BP and GIT methods is roughly same.

In addition, one can find that the main computational time is spent by a lot of calls to the routine computing the values of the integrands. The integrand in \eqref{uWebOrrTrs} is a four-dimensional function of $z$, $p$, $y(\tau)$ and $y(s)$.  Let us denote this functions as $Z(z, p, v, w)$,
where $v = y(\tau)$ and $w = y(s)$. We can gain the speed by the following trick: first pre-compute the values of $Z$ on a regular four-dimensional grid, and then design this computational routine as interpolation.

It is also worth mentioning that in many situations the parameters of the model are such that the boundary $y(\tau)$ changes slowly with time, i.e. $y(\tau)$ is almost const.  Then the first integral in \eqref{uWebOrrTrs} is a good approximation of the price. Accordingly, one don't need to solve the Volterra equation \eqref{VolterraGIT} that makes the algorithm about 2.5 faster.

\section{Connection to physics}

We have already mentioned that both the method of heat potentials (which in our case is extended to the method of Bessel potentials) and the method of generalized integral transforms were first developed in physics, see \citep{kartashov1999, kartashov2001,TS1963, Friedman1964} and references therein. This was done to solve various problems of heat and mass transfer which are widely present in physics, chemistry, energetic, nuclear engineering, geology and many other areas of science and engineering. It turns out that many of those problems mathematically can be formulated in terms of stationary and non-stationary heat transfer. This includes such problems as diffusion, sedimentation, viscosity flows accompanied by various kinetic processes, astronomy, atomic physics, absorbtion, combustion, phase transitions and many others.

As an example in this Section we consider the Stefan problem which is a particular kind of the boundary value problem for the heat equation adapted to the case in which a phase boundary can move with time. It was introduced by I. Stefan in 1889 (see a detailed review in \citep{Lyubov1978}). The classical Stefan problem describes the temperature distribution in a homogeneous medium undergoing a phase change, for example ice passing to water. This is accomplished by solving the heat equation imposing the initial temperature distribution on the whole medium, and a particular boundary condition, the Stefan condition, on the evolving boundary between its two phases. Note that this evolving boundary is an unknown hyper-surface; hence, Stefan problems are examples of free boundary problems. However, a temperature gradient at this boundary is supposed to be known.

As such, treating it in financial terms, one can immediately recognize this as the pricing problem for the American option where the exercise boundary is also a free boundary, i.e. is not known. However, the option Delta $\fp{u}{z}$ at the boundary $z = y(\tau)$ is known (it follows from the conditions $\fp{C_A}{S}|_{S = S_B(t)} = 1$ and $\fp{P_A}{S}|_{S = S_B(t)} = -1$. Also the boundary condition for the American Call and Put at the exercise boundary (the moving boundary) is set as $C_A(S_B(t),t)) = S_B(t) - K$ for the Call, and $P_A(S_B(t),t)) = K - S_B(t)$ for the Put. This problem was solved in \citep{CarrItkin2020} by using the method of generalized integral transform for the time-dependent OU model, and in \citep{LiptonKaush2020} for the Black-Scholes model with constant coefficients by using the method of heat potentials.

On contrary, we want to emphasize that many of the problems considered in this paper have not been solved yet in physics, and are mentioned in \citep{kartashov2001} as yet unsolved problems. Therefore, the results obtained in this paper also make a contribution to physics as they can be easily re-formulated in terms of the above mentioned physics problems.

Another connection of our results with physics is about the first passage time (FPT) problem considered in Section~\ref{FPT}. As mentioned in \citep{Ding2004} (see also references therein), the FPT problem finds applications in many areas of science and engineering . A sampling of these applications includes (but not limited to)
\begin{itemize}
\item statistical physics (study of anomalous diffusion)
\item neuroscience (analysis of neuron firing models)
\item civil and mechanical engineering (analysis of structural failure)
\item chemical physics (study of noise assisted potential barrier crossings)
\item hydrology (optimal design of dams)
\item imaging (study of image blurring due to hand jitter)
\end{itemize}

In particular, in \citep{Redner2001} the author analyses the fundamental connection between the first-passage properties of diffusion and electrostatics. Basic questions of first passage include where a diffusing particle is absorbed on a boundary and when does this absorption event occur. These are time-integrated attributes, obtained by integration of a time-dependent observable over all time. For example, to
determine when a particle is absorbed, we should compute the first-passage probability to the boundary and then integrate over all time to obtain the  eventual hitting probability. However, it is more elegant to reverse the order of calculation and first integrate the equation of motion over time and then compute the outgoing flux at the boundary. This first step transforms the diffusion equation to the simpler Laplace equation. Then,  in computing the flux, the exit probability is just the electric field at the boundary point. Thus, there  is a complete correspondence between a first-passage problem and an electrostatic problem in the same geometry. This mapping is simple yet powerful, and can be adapted to compute related time-integrated properties, such as the splitting probabilities and the moments of the exit time.

Other connections to physics problems such as kinetics of spin systems, first passage in composite and fluctuating systems, hydrodynamic transport. reaction-rate theory, etc. could also be found in \citep{Redner2001}.

With the hope that we managed to convince the reader about a strong connection between the subject of this paper and physics, we stop here this excursus into a wonderful world of physics leaving curious readers to extend it themselves.

\section{Discussion}

In Sections~\ref{SecBO}, \ref{sCIR}  we constructed semi-closed form solutions for the prices of Up-and-Out barrier Call option $C_{uao}$. Despite same could be done for the Down-and-Out options, alternatively we can use the parity for barrier options, \citep{hull97}. Then the price of the Down-and-Out barrier Call option $C_{dao}$ can be found as $C_{dao} = C_{v} - C_{uao}$, where $C_{v}$ is the price of the European vanilla Call option. For the models considered in this paper the latter is known in closed form, \citep{andersen2010interest}. The double barrier case was also considered in Section~\ref{DBO}.

 Despite in our test we assumed the barrier $H$ to be constant in time, the whole framework is developed for the genera case where the barrier is some arbitrary function of time.

From the computational point of view the proposed solution is very efficient as this is shown in Section~\ref{numRes}. Using theoretical analysis justified by a test example we conclude that our method is, at least, of the same complexity, or even faster than the forward FD method. On the other hand, our approach provides high accuracy in computing the options prices, as this is regulated by the order of a quadrature rule used to discretize the kernel. Therefore,  the accuracy of the method in $z$ space can be easily increased by using high order quadratures. On the other hand doing same for the FD method is not easy (i.e., it significantly increases the complexity of the method, e.g., see \citep{ItkinBook}).

Another advantage of the approach advocated in this paper is computation of option Greeks. Since the option prices in both the BP and GIT methods are represented in closed form via integrals, the explicit dependence of prices on the model parameters is available and transparent.
Therefore, explicit representations of the option Greeks can be obtained by a simple differentiation under the integrals. This means that the values of Greeks can be calculated simultaneously with the prices almost with no increase in time. This is because differentiation under the integrals
slightly changes the integrands, and these changes could be represented as changes in weights of the quadrature scheme used to numerically compute the integrals. Since the major computational time has to be spent for computation of densities which contain special functions, they can be saved during the calculation of the prices, and then reused for computation of Greeks.

Note, that the FD method also provides the values of Delta, Gamma and Theta on the FD grid, while, for instance, for Vega one need to bump the model volatility and rerun the whole scheme. But for the BP and GIT methods computation of Delta or Vega is done uniformly. Also, the ability of fast computation of Greeks is important for model calibration. Therefore, one can efficiently calibrate the CIR and CEV models to the market data by using the BP and GIT methods, since the semi-explicit nature of the final expressions allows quasi-analytical formulae for the gradient of the loss function.

\section*{Acknowledgments}

We are grateful to Daniel Duffy and Alex Lipton for some helpful comments. Dmitry Muravey acknowledges support by the Russian Science Foundation under the Grant number 20-68-47030.

\vspace{0.4in}

\vspace{0.4in}
\appendixpage
\appendix
\numberwithin{equation}{section}
\setcounter{equation}{0}

\section{General construction of the potential method.}

In this Section we generalize the construction of the potential method originally proposed for the heat equation. For convenience, we follow the notation of \citep{TS1963}.

Consider a PDE
\begin{equation} \label{gPDE}
\fp{V(t,x)}{t} = \calL(V(t,x))
\end{equation}
\noindent where the operator $\calL$ is a linear differential operator with time-independent coefficients. An example of such the equation is the heat equation and the Bessel equation in \eqref{Bess}. Suppose that the fundamental solution (or the Green function) of \eqref{gPDE}
$G(x,t| \xi,\tau)$ is known in closed form.

Suppose we need to solve \eqref{gPDE} subject to the homogeneous initial condition
\begin{equation} \label{gtc}
V(0,x) = 0.
\end{equation}
Note that this doesn't restrict our consideration, as by the change of variables in  \eqref{q1} the problem with inhomogeneous initial condition can be transformed to the problem with $V(0,x) = 0$.

We are interesting in solving the first boundary-value problem assuming that the boundaries are some functions of time. Consider, for simplicity just
a semi-bounded region with $ x \ge y(t), \ x < \infty, \ t > 0$, with the following boundary conditions
\begin{equation} \label{gbc}
V(t,x)\Big|_{x \to \infty} = 0, \qquad V(t,y(t)) = \phi(t).
\end{equation}
The assumption that the problem has just one time-dependent boundary can be easily relaxed as this is demonstrated in Section~\ref{DBO}, and is not a restriction of the method.

Let us introduce the single layer potential, \citep{TS1963}
\begin{equation} \label{gPoten}
\Pi(x,t) = \int_0^t \Psi(\tau) \fp{G}{\xi}(x,t|\xi,\tau)\Big|_{\xi = y(\tau)} d\tau,
\end{equation}
\noindent where $\Psi(t)$ the potential density. The single layer potential is continuous and twice differentiable function in $x$, and continuous and differentiable in $t$. Then the below proposition follows

\begin{proposition}
The potential function in \eqref{gPoten} solves \eqref{gPDE}.  The potential density is determined by the boundary condition in \eqref{gbc} and solves the Volterra equation of second kind

\begin{equation} \label{gVol}
\phi(t) = b \Psi(t) + \int_0^t \Psi(\tau) \fp{G}{\xi}(y(t),t|y(\tau),\tau)\Big|_{\xi = y(\tau)} d\tau,
\end{equation}
\noindent where $b$ is a constant.

\begin{proof}[{\bf Proof}]
First, it can be checked that substituting the definition in \eqref{gPoten} into \eqref{gPDE} we get
\begin{align}
\fp{\Pi(x,t)}{t} &= \int_0^t \Psi(\tau) \fp{}{\xi} \fp{G}{t}\Big|_{\xi = y(\tau)}  d\tau + \Psi(t) \fp{G}{y(\tau)}(x,y(t)|t,t) d\tau  = \int_0^t \Psi(\tau) \fp{}{\xi} \fp{G}{t}\Big|_{\xi = y(\tau)}  d\tau, \\
\calL(\Pi(t,x))  &= \int_0^t \Psi(\tau) \fp{}{\xi} \calL(G)\Big|_{\xi = y(\tau)} d\tau. \nonumber
\end{align}
Combining these two expressions yields
\begin{equation}
\int_0^t \Psi(\tau) \fp{}{\xi} \left[ \fp{G}{t}   - \calL(G) \right] \Big|_{\xi = y(\tau)} d \tau = 0.
\end{equation}
The first line follows from the fact that the Green function solves \eqref{gPDE}, and $G(x,y(\tau)|t,\tau)\Big|_{\tau = t} = G(x,y(\tau)|t-\tau)\Big|_{\tau = t} = \delta(x-y(t) = 0$ as $x \ne y(t)$. The second line is a consequence of time-independence of the operator coefficients.

To summarize what we got: the potential function satisfies the PDE for $x \ge y(t)$, is bounded at infinity, and has a zero initial value for any choice of $\Psi(t)$. Thus, it is the solution of \eqref{gPDE}, i.e.
\begin{equation} \label{gsol}
V(t,x) = \int_0^t \Psi(\tau) \fp{G}{\xi}(x,t|\xi,\tau)\Big|_{\xi = y(\tau)} d\tau.
\end{equation}
Now using the boundary condition at $x = y(t)$ we obtain from \eqref{gsol}
\begin{equation} \label{gVol1}
\phi(t) = \int_0^t \Psi(\tau) \fp{G}{\xi}(y(t),t|\xi,\tau)\Big|_{\xi = y(\tau)} d\tau.
\end{equation}
However, as shown in \citep{TS1963}, for $x = y(t)$ the RHS is discontinuous, but with the finite limiting value at $x = y(t) + 0$. The limiting value could be represented as
\[ b\Psi(t) + \int_0^t \Psi(\tau) \fp{G}{\xi}(y(t),t|y(\tau),\tau)\Big|_{\xi = y(\tau)} d\tau. \]
The constant $b$ depends on the particular form of the operator $\calL$. In particular, if $\calL$ is a second order parabolic operator with the diffusion coefficient $a$, then $b = 1/(2 a)$, \citep{TS1963}. Thus, for instance, for $a = 1/2$ we have $b=1$.

\end{proof}

\end{proposition}

\end{document}